\documentclass[universe,article,accept,pdftex,moreauthors]{Definitions/mdpi}
\firstpage{1} 
\pubvolume{11}
\issuenum{2}
\articlenumber{47}
\pubyear{2025}
\copyrightyear{2025}
\externaleditor{Mauro D’Onofrio}
\datereceived{23 December 2024 } 
\daterevised{27 January 2025 } 
\dateaccepted{27 January 2025 } 
\datepublished{1 February 2025 } 
\hreflink{https://doi.org/10.3390/\linebreak universe11020047} 

\usepackage{tipa}
\usepackage{xltabular}
\usepackage{xspace}
\usepackage{courier}
\newcommand{\simba}{\textsc{simba}\xspace}
\newcommand{\simbac}{\textsc{simba-c}\xspace}

\makeatletter
\newcommand\betweenscriptsizeandfootnotesize{\@setfontsize\betweenscriptsizeandfootnotesize{8.2}{7}}
\makeatother

\Title{Core to Cosmic Edge: \boldmath{\texttt{SIMBA-C}}'s New Take on Abundance Profiles in the Intragroup Medium at \boldmath{$z=0$}}

\TitleCitation{Core to Cosmic Edge: \texttt{SIMBA-C}'s New Take on Abundance Profiles in the Intragroup Medium at $z=0$}

\Author{Aviv Padawer-Blatt $^{1,}$*\orcidA{}, Zhiwei Shao $^{2}$\orcidB{}, Renier T. Hough $^{3}$\orcidC{}, Douglas Rennehan $^{4}$\orcidD{}, Ruxin Barr\'e $^{1}$\orcidE{}, \mbox{Vida Saeedzadeh $^{5}$\orcidF{}}, Arif Babul $^{1,6,7,}$*\orcidG{}, Romeel Dav\'e $^{6}$\orcidH{}, Chiaki Kobayashi $^{8}$\orcidJ{}, Weiguang Cui $^{9,10}$\orcidI{}, \mbox{Fran\c{c}ois Mernier $^{11,12}$\orcidK{}} and Ghassem Gozaliasl $^{13,14}$\orcidL{}}

\AuthorNames{Aviv Padawer-Blatt, Zhiwei Shao, Renier T. Hough, Douglas Rennehan, Ruxin Barr\'e, Vida Saeedzadeh, Arif Babul, Romeel Dav\'e, Chiaki Kobayashi, Weiguang Cui, Fran\c{c}ois Mernier and Ghassem Gozaliasl}

\AuthorCitation{Padawer-Blatt, A.; Shao, Z.; Hough, R.T.; Rennehan, D.; Barr\'e, R.; Saeedzadeh, V.; Babul, A.; Dav\'e, R.; Kobayashi, C.; Cui, W.; et~al.}

\address{%
$^{1}$ \quad Department of Physics and Astronomy, University of Victoria, Victoria, BC V8P 1A1, Canada; {barreruxin11@uvic.ca}\\
$^{2}$ \quad Department of Astronomy, School of Physics and Astronomy, and Shanghai Key Laboratory for Particle Physics and Cosmology, Shanghai Jiao Tong University, Shanghai 200240, China; {zwshao@sjtu.edu.cn}\\
$^{3}$ \quad Center for Space Research, North-West University, Potchefstroom 2520, South Africa; {25026097@mynwu.ac.za}\\
$^{4}$ \quad Center for Computational Astrophysics, Flatiron Institute, New York, NY 10010, USA; {drennehan@flatironinstitute.org}\\
$^{5}$ \quad Department of Physics \& Astronomy, Johns Hopkins University, Baltimore, MD 21218, USA; {vsaeedz1@jh.edu}\\
$^{6}$ \quad Leverhulme Visiting Prof., Institute for Astronomy, University of Edinburgh, Royal Observatory, Blackford Hill, Edinburgh EH9 3HJ, UK\\
$^{7}$ \quad Infosys Visiting Chair Professor, Department of Physics, Indian Institute of Science, Bangalore 560012, India\\
$^{8}$ \quad Institute for Astronomy, University of Edinburgh, Royal Observatory, Blackford Hill, Edinburgh EH9 3HJ, UK; {romeel.dave@ed.ac.uk}\\
$^{9}$ \quad Centre for Astrophysics Research, Department of Physics, Astronomy and Mathematics, University of Hertfordshire, Hatfield AL10 9AB, UK; {c.kobayashi@herts.ac.uk}\\
$^{10}$\quad Departamento de F\'isica T\'eorica, Universidad Aut\'onoma de Madrid, E-28049 Madrid, Spain; {weiguang.cui@uam.es}\\
$^{11}$\quad Centro de Investigaci\'on Avanzada en F\'isica Fundamental (CIAFF), Universidad Aut\'onoma de Madrid, \mbox{28049 Madrid, Spain}\\
$^{12}$\quad NASA Goddard Space Flight Center, Greenbelt, MD 20771, USA; {fmernier@umd.edu}\\
$^{13}$\quad Department of Astronomy, University of Marlyand, College Park, MD 20742, USA\\
$^{14}$\quad Department of Computer Science, Aalto University, P.O. Box 15400, FI-00 076 Espoo, Finland; {ghassem.gozaliasl@helsinki.fi}\\
$^{15}$\quad Department of Physics, University of Helsinki, P.O. Box 64, FI-00014 Helsinki, Finland\\
}

\corres{Correspondence: {apadawer@uvic.ca} (A.P.-B.); {babul@uvic.ca} (A.B.)}

\abstract{We employ the \simbac cosmological simulation to study the impact of its upgraded chemical enrichment model (\texttt{Chem5}) on the distribution of metals in the intragroup medium (IGrM). We investigate the projected X-ray emission-weighted abundance profiles of key elements over two decades in halo mass ($10^{13} \leq M_{500}/\mathrm{M_\odot} \leq 10^{15}$). Typically, \simbac generates lower-amplitude abundance profiles than \simba with flatter cores, in better agreement with observations. For low-mass groups, both simulations over-enrich the IGrM with Si, S, Ca, and Fe compared to observations, a trend likely related to inadequate modeling of metal dispersal and mixing. We analyze the 3D mass-weighted abundance profiles, concluding that the lower \simbac IGrM abundances are primarily a consequence of fewer metals in the IGrM, driven by reduced metal yields in \texttt{Chem5}, and the removal of the instantaneous recycling of metals approximation employed by \simba. Additionally, an increased IGrM mass in low-mass \simbac groups is likely triggered by changes to the AGN and stellar feedback models. Our study suggests that a more realistic chemical enrichment model broadly improves agreement with observations, but physically motivated sub-grid models for other key processes, like AGN and stellar feedback and turbulent diffusion, are required to realistically reproduce observed group environments.}


\begin{document}






\section{Introduction}
\label{sec:intro}

The majority of galaxies are not born on their own and do not live isolated lives. They often reside and evolve in collections of gravitationally bound groups of galaxies\endnote{Gravitationally bound collections of galaxies reside in halos whose masses span the range from $\sim$$10^{12.5}\,M_\odot$ to $\gtrsim$$10^{15}\,M_\odot$. Historically, lower-mass systems have been referred to as ``groups'' and higher-mass systems as ``clusters''. In this paper, we examine both categories of systems, but, for convenience, we will refer to them collectively as ``groups'', only because the lower-mass systems numerically dominate our simulated samples.}\mbox{~\citep{ekeGalaxyGroups2dF2006,lovisariScalingPropertiesGalaxy2021}, }with their halos contributing the largest proportion of mass to the total mass in the Universe. These environments are filled with hot diffuse gas that permeates the space between galaxies and extends throughout the dark matter halo, known as the intragroup or intracluster medium. Since we are mainly considering groups in this study, we will refer to this simply as the intragroup medium (IGrM). {As a major component of group environments, the IGrM is an important factor in shaping and regulating the evolution of central galaxies~\citep{lebrunRealisticPopulationSimulated2014,ragone-figueroaBCGMassEvolution2018,ragone-figueroaEvolutionRoleMergers2020,mariniVelocityDispersionBrightest2021,martizziBrightestClusterGalaxies2014,remusOuterHalosVery2017,nipotiSpecialGrowthHistory2017,pillepichFirstResultsIllustrisTNG2018,jacksonStellarMassAssembly2020,hendenBaryonContentGroups2020,bassiniDIANOGASimulationsGalaxy2020}}.

The IGrM consists of a significant fraction of the bound baryons in the Universe. Due to its high temperatures, the IGrM emits X-rays, primarily from thermal bremsstrahlung, bound-free emission, and line emission from highly ionized trace elements. It is thus observable with X-ray telescopes, permitting studies that illuminate the formation and evolution of groups, including when and how group galaxies and the IGrM were enriched with metals.

A group's thermal, dynamic, and enrichment history plays a role in shaping its late-time metal distributions. Many observational studies find flat, largely sub-solar gas-phase metallicity profiles in the IGrM outskirts ($R \gtrsim 0.3R_{500}$) of local massive groups and clusters {(e.g.,} ~\citep{wernerUniformMetalDistribution2013,urbanUniformMetallicityOutskirts2017,lovisariNonuniformityGalaxyCluster2019,ghizzardiIronXCOPTracing2021a,sarkarChemicalAbundancesOutskirts2022}). The average amplitudes and slopes for lower-mass groups are not firmly set (e.g.,~\citep{mernierAstrophysicsRadialMetal2017}). Additionally, observations show higher, though generally still sub-solar or solar, enrichment levels in the inner regions of massive groups and clusters, where the central brightest group/cluster galaxies (BGGs/BCGs) typically reside. These core abundances, however, are found to be quite dependent on the thermodynamic state of a group and its cool-core (CC)/non-cool-core (NCC) classification (e.g.,~\citep{ghizzardiIronXCOPTracing2021a}).

The metallicity of the IGrM is directly related to metal production in galaxies via stellar nucleosynthesis, and the subsequent redistribution and mixing through galactic-scale outflows driven by stellar and AGN feedback~\citep{rennehanMixingMatters2021,applebyLowredshiftCircumgalacticMedium2021,saeedzadehCoolGustyChance2023}, ram pressure stripping~\citep{domainkoEnrichmentICMGalaxy2006,saeedzadehCoolGustyChance2023}, and outflow-driven turbulence~\citep{prasadCoolcoreClustersRole2018,rennehanDynamicLocalizedTurbulent2019,bennettResolvingShocksFilaments2020,lochhaasPropertiesSimulatedCircumgalactic2020,rennehanMixingMatters2021} as galaxies form and evolve. The distribution of the IGrM abundances therefore critically relies upon all of these physical processes.

Cosmological-scale galaxy formation simulations are the leading approach in modern astronomy to holistically model such processes in a manner enabling them to reproduce observed galaxy properties. They are made up of elements representing dark matter, gas, stars, and supermassive black holes (SMBHs); numerical implementations of gravity and hydrodynamics for particle interactions, and radiative cooling and heating; and sub-grid models\endnote{These models, also called sub-resolution models, are denoted as such since they model physical processes occurring on spatial, mass, or temporal scales below the resolution threshold of a simulation. They may be phenomenological physically motivated approximations---or even ad hoc solutions---the validity, realism, and effectiveness of which are subject to debate (for an overview, see \citep{winsbergScienceAgeComputer2010}).} for star formation, stellar feedback, chemical enrichment, dust production and destruction, SMBH growth, and AGN feedback, as well as other possibilities~\citep{somervillePhysicalModelsGalaxy2015,vogelsbergerCosmologicalSimulationsGalaxy2020,crainHydrodynamicalSimulationsGalaxy2023}.

The large dynamic range of physical mechanisms contributing to galaxy evolution (e.g., from the accretion disks around SMBHs to the hot atmospheres in groups) and their complex feedback systems (e.g., metal cooling in gas to feed stellar and AGN feedback), however, make it difficult to improve these simulations with regard to their accuracy across a diversity of properties and observables. With that being said, simulations facilitate experimenting with different methods and sub-grid models to address such numerical challenges, continuously refining and reforming them in an effort to better represent galaxies as we observe them.

Sub-grid models and their parameters are regularly constrained by calibrating simulations to selected observables {(e.g.,} ~\citep{vogelsbergerIntroducingIllustrisProject2014,mccarthyBahamasProjectCalibrated2017,pillepichFirstResultsIllustrisTNG2018,daveSimbaCosmologicalSimulations2019,kugelFLAMINGOCalibratingLarge2023,houghSIMBACUpdatedChemical2023a}). For various reasons (including the idea that, by generating realistic simulated galaxies, structures at other scales should effectively be constrained to the correct properties and features), it is rare to employ group-scale observables in this way\endnote{The limited statistics of galaxy groups and clusters compared to individual galaxies in both observations and cosmological simulations additionally play a role in this methodology.}. As a result, the properties of simulated samples of galaxy groups are often discrepant with observations (e.g.,~\citep{vogelsbergerDustGalaxiesDust2019,oppenheimerSimulatingGroupsIntraGroup2021,donahueBaryonCyclesBiggest2022,altamuraEAGLElikeSimulationModels2023,braspenningFLAMINGOProjectGalaxy2024}).

Developing sub-grid models that perform reliably at cosmological resolution, and their continuous testing against observations, is essential for improving the fidelity of simulations, and supports interfacing with observational surveys across an increasingly wide range of wavelengths, scales, and times. Consequently, many studies focus on improving the various sub-grid models by increasing their realism. One such recent step forward is \simbac~\citep{houghSIMBACUpdatedChemical2023a,houghSimbaCEvolutionThermal2024a}. \simbac employs the \simba galaxy formation model~\citep{daveSimbaCosmologicalSimulations2019}, but with an updated stellar feedback and chemical enrichment model, known as \texttt{Chem5}~\citep{kobayashiNewTypeIa2020,kobayashiOriginElementsCarbon2020}. \simba successfully reproduces a wide range of observations while only being tuned to match the $z=0$ galaxy stellar mass function (GSMF) and the $M_\mathrm{BH}-M_*$ relation~\citep{daveSimbaCosmologicalSimulations2019}. \simbac builds on these successes by implementing a more realistic treatment of galactic chemical enrichment. Additionally, \simbac includes refinements to other sub-grid models, such as a reduced stellar wind velocity scaling and delayed activation of AGN jets, to improve agreement with key observations.

At the galaxy level, \simbac performs better than \simba with respect to matching observations~\citep{houghSIMBACUpdatedChemical2023a}. For example, the \simbac $z=0$ GSMF matches observations noticeably better than \simba, but where \simbac shines is in its predictions of metallicity and chemical abundance~\citep{houghSIMBACUpdatedChemical2023a}. Specifically, it yields a significantly improved agreement with observed galaxy scaling relations of $[O/Fe] - [Fe/H]$ and $[Mg/Fe] - [Fe/H]$, and a better match to the $z=0$ stellar mass--metallicity relation at the low-mass end and to the gas-phase $O/H - M_*$ relation.

Hough et~al. \cite{houghSimbaCEvolutionThermal2024a} recently demonstrated that, at $z=0$, compared to \simba, the IGrM of \simbac galaxy groups has higher X-ray luminosity, lower entropy, lower Fe and Si mass-weighted global abundances, and larger Si/O abundance ratios. They also found non-trivial differences in the redshift evolution of the $k_\mathrm{B}T = 1$ keV IGrM metallicity between \simba and \simbac. These results indicate that the inclusion of the updated chemical enrichment model, as well as alterations to the feedback models, has substantial impacts on the thermal properties and metal content of the IGrM.

The \simbac galaxy formation model represents a significant improvement compared to \simba with regard to stellar feedback and chemical enrichment. It is therefore of interest to investigate how the implementation of \texttt{Chem5} in a cosmological simulation and its treatment of physical processes alter the detailed distribution of metals in the IGrM, and if the resulting abundance profiles are in better agreement with observations. Thus, we focus on the radial IGrM metal abundance profiles of \simba and \simbac galaxy groups at $z=0$.

In this article, {Section}~\ref{subsec:methods_sims} reviews the simulation methodology and input physics of \simba~\citep{daveSimbaCosmologicalSimulations2019} and \simbac~\citep{houghSIMBACUpdatedChemical2023a}, highlighting how they differ. Section~\ref{subsec:methods_obs_samples} provides a brief overview of the observational samples used in this work. Sections~\ref{subsec:methods_group_selection}--\ref{subsec:methods_calculating_abundance_profiles} describe the methods employed to obtain and analyze the simulated and observed galaxy groups and their abundance profiles. Section~\ref{sec:results} presents the projected emission-weighted abundance profiles. We compare \simba and \simbac, noting differences and similarities, and compare their profiles to representative observational results, looking at a variety of astrophysically relevant elements. In Section~\ref{sec:discussion}, we assess how much \simbac improves over \simba, and investigate the simulations' 3D mass-weighted abundance profiles to comment on potential reasons for any changes between \simba and \simbac. Finally, we summarize our findings and conclusions in Section~\ref{sec:conclusions}, noting the implications of our results for future simulations, (e.g., new iterations of \simba, like \textsc{kiara}) and future observations of galaxy groups (e.g., the X-Ray Imaging and Spectroscopy Mission (\emph{XRISM}; \citep{teamScienceXrayImaging2020,tashiroXRISMXrayImaging2022}), the Advanced Telescope for High ENergy Astrophysics (\emph{Athena}; \citep{barconsAthenaAdvancedTelescope2012,barretAthenaFirstDeep2013a,barretAthenaSpaceXray2020}), the Advanced X-ray Imaging Satellite \mbox{(\emph{AXIS}; \citep{mushotzkyAdvancedXrayImaging2019a,reynoldsOverviewAdvancedXray2023}),} and \emph{Lynx}~\citep{gaskinLynxXRayObservatory2019,schwartzLynxXrayObservatory2019}).

\section{Methods}
\label{sec:methods}

\subsection{Simulation Methodology}
\label{subsec:methods_sims}

Here, we summarize the \simba and \simbac cosmological hydrodynamic simulations, their physics, and their galaxy formation models and associated sub-grid models. Detailed descriptions can be found in the works of Dav\'e et~al. \cite{daveSimbaCosmologicalSimulations2019} and Hough et~al. \cite{houghSIMBACUpdatedChemical2023a}, respectively.

We employ the flagship \simba\endnote{\url{https://simba.roe.ac.uk/} (accessed on 1 January 2023).} and \simbac runs, each having volumes of side length $100\,\text{h}^{-1}\,\mathrm{cMpc}$ with $1024^3$ dark matter particles and $1024^3$ gas elements. They are run on the same initial conditions from an initial redshift of $z = 249$ down to $z = 0$ assuming a Planck 2018 \cite{planckcollaborationPlanck2018Results2020} flat $\Lambda$CDM cosmology with $\Omega_m = 0.3$, $\Omega_\Lambda = 0.7$, $\Omega_b = 0.048$, and $H_0 = 68 \, \mathrm{km \, s^{-1} \, Mpc^{-1}}$. Both simulations have a minimum Plummer-equivalent gravitational softening length of $0.5\,\text{h}^{-1}\,\mathrm{kpc}$, an initial gas element mass resolution of $1.82 \times 10^7 \, \mathrm{M_{\odot}}$, and a dark matter particle mass resolution of $9.6 \times 10^7 \, \mathrm{M_{\odot}}$. \simba and \simbac are built on the N-body gravity + hydrodynamics solver \textsc{GIZMO} in its meshless finite-mass (MFM) mode~\citep{lansonRenormalizedMeshfreeSchemes2008,lansonRenormalizedMeshfreeSchemes2008a,gaburovAstrophysicalWeightedParticle2011,hopkinsNewClassAccurate2015,hopkinsNewPublicRelease2017}.

In Table~\ref{tab:simba_vs_simbac_sub-grid}, we summarize the key differences between the sub-grid models of \simba and \simbac\endnote{In Hough et~al. \cite{houghSIMBACUpdatedChemical2023a}, the dust model was excluded from \simbac to remove its possible effects on the metal content in a simulation dedicated to testing a new metal enrichment model. Hough et~al. \cite{houghSimbaCEvolutionThermal2024a} re-integrated the dust model into \simbac, finding no discernible effects on the global IGrM properties, likely due to the dust sputtering back into metals in the hot diffuse gas.}. Most importantly, \simbac incorporates an updated chemical enrichment and stellar feedback model (\texttt{Chem5}; \citep{kobayashiOriginElementsCarbon2020,kobayashiNewTypeIa2020}) that replaces the \emph{instantaneous recycling approximation} used in \simba. {This approximation (first described by Talbot and Arnett \cite{talbotEvolutionGalaxiesFormulation1971} and reviewed by Tinsley \cite{tinsleyEvolutionStarsGas2022}) is implemented in \simba as a simple sub-grid model according to the following:}
{
\begin{enumerate}
    \item All gas elements that are eligible to form stars, i.e., the ``star-forming ISM'' (see Dav\'e et~al. \cite{daveMUFASAGalaxyFormation2016} for details), self-enrich with SNII and prompt (concurrent with SNII) SNIa nucleosynthesis products. This occurs at each timestep a gas element is ``star forming'' and has not spawned a star particle.
    \item In the timestep when a star-forming gas element converts to a star particle, no self-enrichment occurs. Also, star particles do not self-enrich.
    \item Feedback from stars below $\sim$$8~\mathrm{M_\odot}$ is delayed relative to the time of star formation. At 0.7 Gyr, after a star particle is spawned, feedback products (metals and energy) from the delayed component of SNe Ia begin to be added to the 16 nearest gas elements in a kernel-weighted manner. Further, AGB stars slowly release---again to the 16 nearest gas elements---these feedback products (as well as mass, with mass conversation maintained by appropriately reducing a star particle's mass) from a star particle over time, beginning at a delay of $\sim$$15$ Myr from the time a star particle spawns.
\end{enumerate}
}

\simbac also employs recalibrated versions~\citep{houghSIMBACUpdatedChemical2023a} of the original AGN and stellar feedback models from \simba. Beyond what is discussed in Table~\ref{tab:simba_vs_simbac_sub-grid}, both simulations employ radiative cooling and photoionization heating from the \textsc{grackle-3.1} library~\citep{smithGrackleChemistryCooling2017}, and an H$_2$-based star formation model with H$_2$ fractions computed according to the method of Krumholz and Gnedin \cite{krumholzCOMPARISONMETHODSDETERMINING2011}.

\subsection{Observational Samples}
\label{subsec:methods_obs_samples}

In this section, we describe the samples of X-ray observational data collected from the literature. This is by no means a comprehensive or exhaustive compilation, but it contains relatively recent results and provides an illustrative representation of the spread across the literature, selection effects due to variations in observing strategy, telescopes and instruments, analysis methods and codes, spectral fitting models, and group-to-cluster scales. Table~\ref{tab:obs_samples} summarizes the key details of each sample.

\begin{table}[H]
    \caption{Summary of key differences between sub-grid models of \simba and \simbac.}
    \label{tab:simba_vs_simbac_sub-grid}
   
    \begin{adjustwidth}{-\extralength}{0cm}
    {\footnotesize
    \begin{tabular}{p{0.17\paperwidth}p{0.32\paperwidth}p{0.32\paperwidth}}
        \toprule
        {\textbf{Sub-Grid Model}} & {\textbf{\simba}} & {\textbf{\simbac}} \\
        \midrule
 &&\\   [-5.ex]
        \vspace{5pt}
        {Star Formation} &
        \setlength{\leftmargini}{7pt}
        \begin{itemize}[labelindent=0cm,labelsep=0.1cm,leftmargin=*]
            \item $\mathrm{SFR}=\varepsilon_* \rho_{\mathrm{H_2}}/t_{\mathrm{dyn}}$ with $\varepsilon_*=0.02$~\citep{kennicuttGlobalSchmidtLaw1998}.
        \end{itemize}
        & 
        \setlength{\leftmargini}{7pt}
        \begin{itemize}[labelindent=0cm,labelsep=0.1cm,leftmargin=*]
            \item Updated value of $\varepsilon_*=0.026$~\citep{pokhrelSinglecloudStarFormation2021}.
        \end{itemize} \\
     &&\\   [-5.ex]
        \midrule
&&\\   [-5.ex]
        \vspace{100pt}
        {Chemical Enrichment} & 
        \setlength{\leftmargini}{7pt}
        \begin{itemize}[labelindent=0cm,labelsep=0.1cm,leftmargin=*]
            \item Elements: H, He, C, N, O, Ne, Mg, Si, S, Ca, Fe.
            \item Metal production by SNe II, SNe Ia, and AGB stars.
            \item Instantaneous recycling of metals model, whereby metals are self-injected into gas elements in the ISM at every timestep they are eligible to form star particles (i.e., above the SF density threshold, where this conversion is performed stochastically), under the assumption that SNe II and the prompt (concurrent with SNe II) SNIa components occur nearly instantaneously after star formation, with the metals being rapidly recycled back into the warm ISM~\citep{daveMUFASAGalaxyFormation2016}.
            \item Delayed feedback component of 0.7 Gyr (relative to time of conversion of gas element into star particle) for AGB stars and SNe Ia, adding mass, metals, and energy to the 16 nearest neighbors in a kernel-weighted manner~\citep{daveMUFASAGalaxyFormation2016}.
            \item Yield tables from Nomoto et~al. \cite{nomotoNucleosynthesisYieldsCorecollapse2006} for SNe II, Iwamoto et~al. \cite{iwamotoNucleosynthesisChandrasekharMass1999} for SNe Ia, and Oppenheimer and Dav\'e \cite{oppenheimerCosmologicalSimulationsIntergalactic2006} for AGBs.
        \end{itemize}
        &
        \setlength{\leftmargini}{7pt}
        \begin{itemize}[labelindent=0cm,labelsep=0.1cm,leftmargin=*]
            \item Adopts \texttt{Chem5} cosmic chemical enrichment model (\citep{kobayashiOriginElementsCarbon2020,kobayashiNewTypeIa2020} and references therein).
            \item Elements: all elements from H to Ge.
            \item Metal production and energetic feedback by core-collapse SNe (SNe II, hypernovae (HNe; high-mass core-collapse SNe whose explosion energy is $>$$10\times$ that of a regular SN), and `failed' SNe for most massive stars), SNe Ia, and stellar winds from stars of all masses (including AGB and super-AGB stars).
            \item No instantaneous recycling of metals or delayed feedback models; enrichment occurs only \emph{{after}} gas element spawns star particle.
            \item Treats each star particle as an evolving stellar population that ejects energy, mass, and metals in a time-resolved sense and distributes these ejecta to the 64 neighboring gas particles in a kernel-weighted fashion, with new analytical models for each enrichment channel (see \citep{kobayashiOriginElementsCarbon2020,kobayashiNewTypeIa2020}).
            \item Updated yield tables (with generally reduced yields) for each enrichment channel. {See reference \cite{kobayashiGalacticChemicalEvolution2006} for SNe II/HNe, \cite{kobayashiOriginElementsCarbon2020} for ‘failed’ SNe, \cite{kobayashiNewTypeIa2020} for SNe Ia, and \cite{dohertySuperMassiveAGB2014a,dohertySuperMassiveAGB2014} for AGBs/super AGBs.}
        \end{itemize} \\
&&\\   [-5.ex]
        \midrule
&&\\   [-5.ex]
        \vspace{110pt}
        {Stellar Feedback} & 
        \setlength{\leftmargini}{7pt}
        \begin{itemize}[labelindent=0cm,labelsep=0.1cm,leftmargin=*]
            \item Gas elements eligible for SF can self-kick into a wind, with probability equal to the probability of converting into a star times the mass-loading factor $\eta$ (as calculated by Dav\'e et~al.~\citep{daveSimbaCosmologicalSimulations2019}), assuming massive stars launch two-phase kinetic winds that drive material out of galaxies through SNe II, radiation pressure, and stellar winds~\citep{daveMUFASAGalaxyFormation2016}.
            \item Stellar mass loss due to SNe II of $f_{\mathrm{SNII}}=0.18$ of each gas element's mass, which is released to the warm ISM at each timestep it is eligible for SF.
            \item Wind velocity dependent on $v_{\mathrm{circ}}$, based on the study of Muratov et~al. \cite{muratovGustyGaseousFlows2015}, with scaling of $a=1.6$.
            \item Of the stellar wind particles, 30\% are ejected ``hot'' and hydrodynamically decoupled for a short time.
            \item Ejected winds are metal loaded by extracting some metals from nearby particles to represent local enrichment by SNe, with the metallicity added given by $dZ = f_{\mathrm{SNII}}y_{\mathrm{SNII}}(Z)/\mathrm{MAX}(\eta,1)$.
            \item \textls[-15]{Long-lived stars provide energetic feedback through SNIa and AGB delayed components by dumping energy and mass into the 16 nearest neighbors in a kernel-weighted manner, heating these gas elements under the assumption that the energy thermalizes~rapidly.}
        \end{itemize}
        & 
        \setlength{\leftmargini}{7pt}
        \begin{itemize}[labelindent=0cm,labelsep=0.1cm,leftmargin=*]
            \item Same two-phase decoupled kinetic wind model, with some changes, as follows.
            \item $f_{\mathrm{SNII}}=0$ due to the removal of the instantaneous recycling approximation.
            \item The assumption that SNe II provide energy for kicking gas elements into stellar winds remains; however, all other stellar channels of energetic feedback are now computed by \texttt{Chem5}, with the energy and mass only released in the timesteps after a star particle has formed.
            \item Stellar wind velocity scaling reduced to $a=0.85$ to match the median value of Muratov et~al. \cite{muratovGustyGaseousFlows2015}.
            \item The fraction of wind particles ejected hot and hydrodynamically decoupled follows the trend from \textsc{FIRE} simulations~\citep{pandyaUnifiedModelCoevolution2023}.
            \item Winds are not artificially metal loaded, since they naturally contain metals in \texttt{Chem5}.
        \end{itemize} \\
&&\\   [-5.ex]
        \midrule
&&\\   [-5.ex]
        \vspace{14pt}
        {Black Hole Formation} &
        \setlength{\leftmargini}{7pt}
        \begin{itemize}[labelindent=0cm,labelsep=0.1cm,leftmargin=*]
            \item BHs are seeded in galaxies not already containing a BH that reach a stellar mass $M_* \gtrsim 3 \times 10^{9} \, \mathrm{M_\odot}$, with $M_\mathrm{BH,seed}=10^4\,\text{h}^{-1}\,\mathrm{M_\odot}$.
        \end{itemize}
        &
        \setlength{\leftmargini}{7pt}
        \begin{itemize}[labelindent=0cm,labelsep=0.1cm,leftmargin=*]
            \item BHs are seeded as soon as galaxies reach a resolved stellar mass of $M_* \gtrsim 6 \times 10^8 \, \mathrm{M_\odot}$, still with $M_\mathrm{BH,seed}=10^4\,\text{h}^{-1}\,\mathrm{M_\odot}$.
        \end{itemize} \\
&&\\   [-5.ex]
        \midrule
&&\\   [-5.ex]
        \vspace{20pt}
        {Black Hole Growth} & 
        \setlength{\leftmargini}{7pt}
        \begin{itemize}[labelindent=0cm,labelsep=0.1cm,leftmargin=*]
            \item The BH mass accretion rate is a linear combination of two modes as $\dot{M}_{\mathrm{BH}}=(1-\eta) \times (\dot{M}_{\mathrm{Torque}} + \dot{M}_{\mathrm{Bondi}})$ for each BH's ``life'', with $\eta=0.1$.
        \end{itemize}
        & 
        \setlength{\leftmargini}{7pt}
        \begin{itemize}[labelindent=0cm,labelsep=0.1cm,leftmargin=*]
            \item BH accretion is the same as in \simba, but with the initial accretion rate (for $M_{\mathrm{BH}} < 3 \times 10^6 \mathrm{M_\odot}$) suppressed by a factor of $e^{-M_{\mathrm{BH}}/10^6 \, \mathrm{M_\odot}}$, to simulate star formation suppressing BH growth in dwarf galaxies.
        \end{itemize} \\
&&\\   [-5.ex]
\midrule
&&\\   [-5.ex]
        \vspace{50pt} {Black Hole Feedback} &
        \setlength{\leftmargini}{7pt}
        \begin{itemize}[labelindent=0cm,labelsep=0.1cm,leftmargin=*]
            \item Two modes: X-ray heating and bipolar kinetic outflows, with the latter divided into jets (only at $f_{\mathrm{Edd}}<0.2$) and radiative winds (see \citep{daveSimbaCosmologicalSimulations2019} for  details).
            \item Jets activate between $M_{\mathrm{BH,jet,min}}=4 \times 10^7 \, \mathrm{M_\odot}$ and $M_{\mathrm{BH,jet,max}}=6 \times 10^7 \, \mathrm{M_\odot}$, with probability scaling from $0 \rightarrow 1$ over that mass range.
            \item Jet velocity increases with decreasing $f_{\mathrm{Edd}}$ as $v_{\mathrm{jet}} = 7000 \, \mathrm{km \, s^{-1}} \times \log_{10}\big(0.2/\mathrm{MAX}(f_{\mathrm{Edd}}, 0.02)\big)$ until $f_{\mathrm{Edd}}=0.02$, below which it is capped at $7000 \, \mathrm{km \, s^{-1}}$.
        \end{itemize}
        &
        \setlength{\leftmargini}{7pt}
        \begin{itemize}[labelindent=0cm,labelsep=0.1cm,leftmargin=*]
            \item Same models, with alterations, as follows.
            \item Jet activation range increased to $M_{\mathrm{BH,jet,min}}=7 \times 10^7 \, \mathrm{M_\odot}$ and $M_{\mathrm{BH,jet,max}}= 10^8 \, \mathrm{M_\odot}$, with specific jet onset mass assigned to each BH particle.
            \item Maximum jet velocity additionally scales with $M_{\mathrm{BH}}$ at $f_{\mathrm{Edd}}<0.02$ as $v_{\mathrm{jet}} = 7000 \, \mathrm{km \, s^{-1}} \, \times \mathrm{MIN}\big(\log_{10}(0.2/f_{\mathrm{Edd}}), v_{\mathrm{max}}\big)$, where $v_{\mathrm{max}} = \big(M_{\mathrm{BH}}/(10^8 \mathrm{M_\odot})\big)^{1/3}$ is limited to the range $[1,5]$, such that $v_\mathrm{jet}$ is capped at $35,000 \, \mathrm{km \, s^{-1}}$ for $M_\mathrm{BH} \geq 1.25 \times 10^{10} \, \mathrm{M_\odot}$ and $f_\mathrm{Edd} \leq 2 \times 10^{-6}$.
        \end{itemize} \\
&&\\   [-5.ex]
        \bottomrule
    \end{tabular}
    }
    \end{adjustwidth}
\end{table}

Werner et~al. \cite{wernerXMMNewtonHighresolutionSpectroscopy2006} measure core IGrM abundance profiles in the giant elliptical galaxy M87 (near/at the Virgo Cluster center) using high-resolution spectra obtained with the Reflection Grating Spectrometers (RGS; \citep{denherderReflectionGratingSpectrometer2001}) on board \emph{{XMM-Newton}}~\citep{jansenXMMNewtonObservatorySpacecraft2001}.

Grange et~al. \cite{grangeMetalContentsTwo2011} employ the RGS and EPIC (European Photon Imaging \mbox{Camera;~\citep{struderEuropeanPhotonImaging2001,turnerEuropeanPhotonImaging2001})} instruments on board \emph{XMM-Newton} to measure core IGrM abundances in two galaxy groups, NGC 5044 and NGC 5813.

The CHEmical Evolution RGS Sample (\textsc{CHEERS}) consists of 44 nearby bright cool-core galaxy clusters, groups, and ellipticals observed with \emph{XMM-Newton}~\citep{deplaaCHEERSChemicalEvolution2017}. Mernier et~al. \cite{mernierAstrophysicsRadialMetal2017} perform spectral fitting with the EPIC instrument and construct IGrM abundance profiles centered on the objects' X-ray emission peaks.

Mao et~al. \cite{maoNitrogenAbundanceXray2019} perform spectral fitting of both \emph{XMM-Newton}/RGS and EPIC/MOS instruments to determine core IGrM abundances in \textsc{CHEERS}. The nitrogen abundances could only be well constrained in eight groups, resulting in a smaller sample than \textsc{CHEERS}.

The XMM Cluster Outskirts Project (\textsc{X-COP}) consists of observations of the outer regions of 13 massive clusters ($3 \times 10^{14} < M_{500}/\mathrm{M_\odot} < 9 \times 10^{14}$) at $z=0.04-0.1$~\citep{eckertXMMClusterOutskirts2017}. For objects in X-COP, Ghizzardi et~al.~\citep{ghizzardiIronXCOPTracing2021a} carry out joint spectral fitting with the \emph{XMM-Newton} EPIC/MOS and pn instruments.

Fukushima et~al. \cite{fukushimaNeMgFe2023} measure the core IGrM abundances of 14 \textsc{CHEERS} groups with \emph{XMM-Newton}/RGS.

Lastly, Sarkar et~al. \cite{sarkarChemicalAbundancesOutskirts2022} investigate chemical abundances out to $\sim$$R_{200}$ in the IGrM of four nearby galaxy groups with the \emph{Suzaku}~\citep{mitsudaXRayObservatorySuzaku2007} X-Ray Imaging Spectrometer (XIS; \citep{koyamaXRayImagingSpectrometer2007}) (for spectral fitting) and \emph{Chandra} \citep[][]{weisskopfChandraXrayObservatory2000} (for pinning down uncertainty introduced by the Cosmic X-Ray Background (CXB)).

\vspace{-3pt}
\begin{table}[H]
\setlength{\tabcolsep}{3.5mm}
    \caption{{Selected observational samples containing intragroup medium (IGrM) abundance profiles. Note that telescopes are written in \emph{italics}, while samples and instruments are in capitalized Roman~font.}}
    \label{tab:obs_samples}
    \begin{adjustwidth}{-\extralength}{0cm}
    {\footnotesize
      \begin{tabular}{ccccccc}
        \toprule
        \multirow{2}{*}{\textbf{Paper/Sample}}& \vtop{\hbox{\strut \textbf{\hspace{2mm}Telescope}}\hbox{\strut \textbf{Instrument(s)}}}& \vtop{\hbox{\strut \textbf{X-Ray Band}}\hbox{\strut \textbf{\hspace{5mm}[keV]}}}& \vtop{\hbox{\strut \textbf{\hspace{1mm}No. Objects/}}\hbox{\strut \textbf{Object Names}}}& \vtop{\hbox{\strut \textbf{Spatial Extent}}\hbox{\strut \textbf{(Max. Radius)}}}& \multirow{2}{*}{\textbf{Elements}}\\
       \midrule
         \rule{0pt}{2.5ex} \multirow{2}{*}{Werner et~al., 2006 \cite{wernerXMMNewtonHighresolutionSpectroscopy2006}}& \vtop{\hbox{\strut \emph{XMM-Newton}}\hbox{\strut \hspace{5mm}RGS}}& \multirow{2}{*}{0.8{--}1.4} & \multirow{2}{*}{M87}& \multirow{2}{*}{$\sim$$0.02R_{500}$}& \multirow{2}{*}{C, N, O, Ne, Fe}\\[3ex]
         \midrule
         \rule{0pt}{2.5ex} \multirow{2}{*}{Grange et~al., 2011 \cite{grangeMetalContentsTwo2011}}& \vtop{\hbox{\strut \emph{\hspace{5mm}XMM-Newton}}\hbox{\strut RGS, EPIC/MOS+pn}}&  \multirow{2}{*}{$\sim$0.35--10}&  \vtop{\hbox{\strut NGC 5044}\hbox{\strut NGC 5813}}& \vtop{\hbox{\strut $\sim$$0.06R_{500}$}\hbox{\strut $\sim$$0.08R_{500}$}}& \vtop{\hbox{\strut C, N, O, Ne, Mg, Si,}\hbox{\strut S, Ar, Ca, Fe, Ni}}\\[3ex]
         \midrule
         \rule{0pt}{2.5ex} \vtop{\hbox{\strut Mernier et~al., 2017 \cite{mernierAstrophysicsRadialMetal2017}}\hbox{\strut \hspace{6mm}(\textsc{CHEERS})}}& \vtop{\hbox{\strut \emph{\hspace{1mm}XMM-Newton}}\hbox{\strut EPIC/MOS+pn}}& \multirow{2}{*}{0.5--10}& \multirow{2}{*}{44}& \vtop{\hbox{\strut $\sim$$0.6R_{500}$ (Groups)}\hbox{\strut $\sim$$0.9R_{500}$ (Clusters)}}& \vtop{\hbox{\strut O, Mg, Si, S,}\hbox{\strut Ar, Ca, Fe, Ni}}\\[3ex]
         \midrule
         \rule{0pt}{2.5ex} \multirow{2}{*}{Mao et~al., 2019 \cite{maoNitrogenAbundanceXray2019}}& \vtop{\hbox{\strut \emph{\hspace{2mm}XMM-Newton}}\hbox{\strut RGS, EPIC/MOS}}& \multirow{2}{*}{$\sim$0.5--1.8}& \multirow{2}{*}{8}& \multirow{2}{*}{$\sim$0.01--0.04$R_{500}$}& \multirow{2}{*}{N, O, Ne, Mg, Fe, Ni}\\[3ex]
         \midrule
         \rule{0pt}{2.5ex} \vtop{\hbox{\strut Ghizzardi et~al., 2021 \cite{ghizzardiIronXCOPTracing2021a}}\hbox{\strut \hspace{10mm}(\textsc{X-COP})}}& \vtop{\hbox{\strut \emph{\hspace{1mm}XMM-Newton}}\hbox{\strut EPIC/MOS+pn}}& \multirow{2}{*}{0.5--12}& \multirow{2}{*}{13}& \multirow{2}{*}{$\sim$$R_{500}$}& \multirow{2}{*}{Fe}\\[3ex]
         \midrule
         \rule{0pt}{2.5ex} \multirow{2}{*}{Fukushima et~al., 2023 \cite{fukushimaNeMgFe2023}}& \vtop{\hbox{\strut \emph{XMM-Newton}}\hbox{\strut \hspace{5mm}RGS}}& \multirow{2}{*}{$\sim$0.5--1.8}& \multirow{2}{*}{14}& \multirow{2}{*}{$\sim$0.03--0.07$R_{500}$}& \multirow{2}{*}{N, O, Ne, Mg, Fe, Ni}\\[3ex]
         \midrule
         \rule{0pt}{2.5ex} \multirow{2}{*}{Sarkar et~al., 2022 \cite{sarkarChemicalAbundancesOutskirts2022}}& \vtop{\hbox{\strut \emph{Suzaku}}\hbox{\strut \hspace{2mm}XIS}}& \multirow{2}{*}{0.5--7}& \multirow{2}{*}{4}& \multirow{2}{*}{$\sim$$2R_{500}$}& \multirow{2}{*}{O, Mg, Si, S, Fe, Ni}\\[3ex]
         \bottomrule
    \end{tabular}
    }
    \end{adjustwidth}
\end{table}

\subsection{Group Selection}
\label{subsec:methods_group_selection}

We identify group halos at $z=0$ as per the method laid out by Jung et~al. \cite{jungMassiveCentralGalaxies2022} and Hough et~al. \cite{houghSimbaCEvolutionThermal2024a}. The AMIGA Halo Finder (AHF; \citep{knebeRelationRadialAlignment2008,knollmannAHFAmigaHalo2009}) is employed to hierarchically identify halos and subhalos, and we find their centers with the shrinking-sphere approach~\citep{powerInnerStructureLCDM2003}.

We characterize the halos by their halo mass $M_\Delta$, the mass within a sphere of radius $R_\Delta$ centered on a halo's center, such that the mean interior mass density is $\Delta$ times the critical density of the Universe. We focus on the local universe ($z=0$), and use $\Delta = 500$ for the characteristic halo masses and radii. The Python package XIGrM\endnote{\url{https://xigrm.readthedocs.io/en/latest/} (accessed on 1 January 2023).} (X-ray properties of the IntraGroup Medium) is utilized to calculate the various X-ray halo quantities.

We select simulated galaxy groups as halos with three or more ``luminous'' galaxies ($N_{\mathrm{l,gal}}$; \citep{houghSimbaCEvolutionThermal2024a}), which are defined as having a stellar mass $M_\star \geq 1.16 \times 10^9 \mathrm{M_\odot}$, equivalent to $\geq 64$ star particles. We also restrict the halo mass range to $M_{500}/\mathrm{M_\odot} \geq 10^{13}$, as the majority of observed groups with measured abundance distributions do not fall below this value. In total, there are 238 halos matching these conditions in \simba, and 258 halos in \simbac. There is a negligible number of halos with $M_{500}/\mathrm{M_\odot} \geq 10^{13}$ and $N_{\mathrm{l,gal}} < 3$, indicating that the cut on $N_{l,\mathrm{gal}}$ has little impact on the results.

For plotting purposes, we derive $R_{2500}$ and $R_{200}$ from $R_{500}$ using approximate conversion factors from the ratios of the median values of our \simbac group sample, with $\overline{R}_{2500} \approx 182.4 \, \mathrm{kpc}$, $\overline{R}_{500} \approx 414.1 \, \mathrm{kpc}$, and $\overline{R}_{200} \approx 641.2 \, \mathrm{kpc}$\endnote{Computing the median values of $R_{500}$ separately in our three $M_{500}$ bins results in negligible changes to their ratios compared to using the whole sample.}.

To investigate how the overall statistical properties of abundance profiles change with group mass, we use $M_{500}$ rather than $M_{200}$ or $M_\mathrm{vir}$, as it is more easily derived from observations of dim groups, and thus better facilitates comparison between simulations and observations. We separate both simulated and observed galaxy groups into three bins in $M_{500}$: $13 < \log(M_{500}/\mathrm{M_\odot}) \leq 13.5$, $13.5 < \log(M_{500}/\mathrm{M_\odot}) \leq 14$, and $14 < \log(M_{500}/\mathrm{M_\odot}) \leq 15$. See Table~\ref{tab:num_groups} for the number of simulated halos in each mass bin.

\begin{table}[H]
\setlength{\tabcolsep}{1.83mm}
    \caption{Number of galaxy groups in \simba and \simbac for the whole sample, and separately for each mass bin.}
    \label{tab:num_groups}
    \begin{adjustwidth}{-\extralength}{0cm}
    \begin{tabular}{ccccc}
        \toprule
         \multirow{2}{*}{\textbf{Simulation}}&  \multirow{2}{*}{\textbf{All Bins}}&  \vtop{\hbox{\strut \boldmath{$13 < \log(M_{500}/\mathrm{M_\odot}) \leq 13.5$}}\hbox{\strut \boldmath{$0.7 \lesssim k_\mathrm{B}T_{\mathrm{spec,corr}} \, [\mathrm{keV}] \lesssim 1.2$}}}&  \vtop{\hbox{\strut \boldmath{$13.5 < \log(M_{500}/\mathrm{M_\odot}) \leq 14$}}\hbox{\strut \boldmath{$1.2 \lesssim k_\mathrm{B}T_{\mathrm{spec,corr}} \, [\mathrm{keV}] \lesssim 2.2$}}}& \vtop{\hbox{\strut \boldmath{$14 < \log(M_{500}/\mathrm{M_\odot}) \leq 15$}}\hbox{\strut \boldmath{$2.2 \lesssim k_\mathrm{B}T_{\mathrm{spec,corr}} \, [\mathrm{keV}] \lesssim 7.4$}}}\\
       \midrule
         \simba&  238&  175&  54& 9\\
         \simbac&  258&  182&  67& 9\\
       \bottomrule
    \end{tabular}
    \end{adjustwidth}
\end{table}

\subsection{IGrM X-Ray Properties}
\label{subsec:methods_xray_properties}

To compute IGrM properties that would observationally be derived from X-ray luminosity or spectral measurements, we follow the procedure of Hough et~al. \cite{houghSimbaCEvolutionThermal2024a}. We define the simulated IGrM as all gas particles with $T > 5 \times 10^5$ K and a hydrogen number density $n_H < 0.13 \,\, \mathrm{atoms \,\, cm^{-3}}$ (below the star formation density threshold in \simba and \simbac). Our tests indicate that the derived abundance profiles are relatively robust to the choice of temperature threshold, from $\sim$$10^5${--}$5 \times 10^{6}$ K.

Both \simba and \simbac launch hydrodynamically decoupled interstellar particles to represent the heated component of stellar and AGN winds. We exclude these wind particles while they are decoupled in all X-ray analyses in the present study.

\subsubsection{X-Ray Luminosity}
\label{subsubsec:methods_xray_luminosity}

We calculate the X-ray luminosity of each gas particle using two Python packages. XIGrM computes the temperature of each gas particle based on its internal energy, hydrogen mass fraction, and electron abundance. \textsc{PyAtomDB}\endnote{The \textsc{PyAtomDB} 
documentation can be found at \url{https://atomdb.readthedocs.io/en/master/} (accessed on 1 January 2023).}~\citep{fosterAtomDBPyAtomDBAtomic2016,fosterPyAtomDBExtendingAtomDB2020,fosterAtomDBPyAtomDBTools2021} generates each gas particle's X-ray spectrum using their mass, temperature, metallicity, and SPH-weighted\endnote{SPH = smoothed particle hydrodynamics.} density under the assumption that the gas is optically thin and in collisional ionization equilibrium.

To obtain the total X-ray luminosity for a gas particle in a desired energy band $L_{X,E_1-E_2,i}$, where $i$ denotes the particle's id, the photon energy intensities are summed across the specified energy range. We use the total summed line and continuum emission to compute the X-ray luminosities. Additionally, we opt to use the 0.5{--}10 keV band, as Mernier et~al. \cite{mernierAstrophysicsRadialMetal2017} use this energy range for spectroscopic fitting. We find negligible differences in the simulated abundance profiles when using other bands.

\subsubsection{Abundances}
\label{subsubsec:methods_abundances}

Observationally, gas-phase abundances in the IGrM are primarily derived from spectroscopic fitting (see \citep{gastaldelloMetalContentHot2021,mernierChemicalEnrichmentGroups2022} and references therein). Gastaldello et~al. \cite{gastaldelloMetalContentHot2021} note the importance of carefully choosing the weight \emph{w} when computing the weighted average metallicity $Z_w = \int w Z dV / \int w dV$ in simulated halos and comparing it to observations.

For comparison to observations, we opt to calculate 2D projected emission-weighted abundances (Equation \eqref{eqn:ZLx}). We describe the projecting procedure in Section~\ref{subsec:methods_calculating_abundance_profiles}. Weighting abundance by X-ray luminosity more closely resembles observations than mass or volume weighting, as the spectroscopically inferred abundances are influenced most strongly by bright areas in a group~\citep{biffiHistoryChemicalEnrichment2017}.

To compute an element's abundance, we calculate an X-ray luminosity weighted mean:
\begin{linenomath}
\begin{equation}
    \label{eqn:ZLx}
    Z_{L_X,q} = \frac{\sum_i Z_{q,i} L_{X,i}}{\sum_i L_{X,i}},
\end{equation}
\end{linenomath}
where $i$ refers to an individual gas particle's id, $q$ is the element under consideration, $Z_{q,i}=m_{q,i}/m_{gas,i}$ is the mass fraction of element $q$ for particle $i$, $L_{X,i}$ is the X-ray luminosity calculated as described previously, and the sum runs over all particles within a halo or radial bin. We scale all abundances to the photospheric solar abundances of Asplund et~al. \cite{asplundChemicalCompositionSun2009}, also rescaling observational results that are reported with different solar normalizations.

One can ask whether emission-weighted abundances are comparable to spectroscopic values. Some studies have found little difference between intrinsic abundance profiles and those inferred from mock spectroscopy of simulated groups (e.g.,~\citep{pearceRedshiftEvolutionHot2021}). Therefore, while it would make for a more realistic comparison, we do not perform full spectral fitting of post-processed synthetic X-ray spectra of galaxy groups (see \citep{jenningsHaloScalingRelations2023b,cuiHYENASProjectPrediction2024a,jenningsHyenasXrayBubbles2025} for current work).

{We emphasize that emission-weighted abundances must be interpreted with caution in the context of metal formation mechanisms and timescales. Hough et~al. \cite{houghSimbaCEvolutionThermal2024a} show the differences in global mass-weighted and emission-weighted abundances, and demonstrate the bias in measuring abundances only from the X-ray-emitting gas. Consequently, when we compare the abundance profiles of \simba and \simbac to investigate the underlying physical processes in Section~\ref{subsec:discussion_changes}, we use 3D spherically averaged mass-weighted profiles, as they better represent the intrinsic abundance distributions. These abundances are calculated~as}
\begin{linenomath}
\begin{equation}
    \label{eqn:Zmass}
    Z_{m,q} = \frac{\sum_i Z_{q,i} m_i}{\sum_i m_i} = \frac{M_q}{M_{\mathrm{gas}}},
\end{equation}
\end{linenomath}
where $M_q$ is the total mass of element $q$ in a halo or radial bin, and $M_{\mathrm{gas}}$ is the total mass of IGrM gas in the same halo or radial bin.

\subsubsection{Global Temperature}
\label{subsubsec:methods_global_temp}

In addition to $M_{500}$, we use an observationally driven global measurement of group temperature to characterize each group. De Plaa et~al. \cite{deplaaCHEERSChemicalEvolution2017} take group temperatures from Chen et~al. \cite{chenStatisticsXrayObservables2007} and Snowden et~al. \cite{snowdenCatalogGalaxyClusters2008} for \textsc{CHEERS} objects. As in these studies, IGrM temperatures are typically determined by identifying a single-temperature thermal model whose spectrum best matches the observed spectrum. Reproducing this procedure is involved and time consuming because the observed spectrum is a combined output of emission from gas that spans a range of temperatures. We instead follow the approach adopted by Hough et~al. \cite{houghSimbaCEvolutionThermal2024a}, and use the temperature measure proposed by Vikhlinin~\cite{vikhlininPredictingSingleTemperatureFit2006}, which we will refer to as the ``spectroscopic'' temperature $T_{\mathrm{spec}}$. Vikhlinin~\cite{vikhlininPredictingSingleTemperatureFit2006} shows that $T_{\mathrm{spec}}$ is a good estimate of the temperature derived from fitting the spectrum.

{Moreover, the cores of groups and clusters are known to have widely varying temperatures that may not correlate well with their total mass or gravitational potential well depth. Observationally, O'Sullivan et~al.~\cite{osullivanCompleteLocalVolume2017} show that the central regions at $R \lesssim 0.15R_{500}$ of both CC and NCC groups exhibit temperature drops or increases, while the temperature profiles outside this core region tend to remain flat to the measured radial extents. \mbox{Jennings et~al.~\cite{jenningsHyenasXrayBubbles2025}} demonstrate that X-ray AGN bubbles in the \textsc{Hyenas}~\citep{cuiHYENASProjectPrediction2024a} suite of zoom-in cosmological simulations typically sit in the IGrM at $\sim$20{--}60 kpc, which roughly corresponds to $\sim$0.05$-$0.2$R_{500}$, agreeing with the extent of the observations. Many observational studies, such as those of Chen et~al.~\cite{chenStatisticsXrayObservables2007} and Snowden et~al. \cite{snowdenCatalogGalaxyClusters2008}, use different methods to excise this inner region from groups with cooling cores when determining group properties. As our primary goal is to sample the gas temperature that is reflective of the gravitational potential, we remove all IGrM particles within $0.15R_{500}$ when computing $T_{\mathrm{spec}}$, restricting ourselves to the radial range $0.15 \leq R/R_{500} \leq 1$. We refer to this as the ``core-corrected'' temperature, or $T_{\mathrm{spec,corr}}$, which is found to be more robust than $T_{\mathrm{spec}}$ (see Hough et~al.~\citep{houghSimbaCEvolutionThermal2024a} for more details).}

We use the $M_{500}-T_{\mathrm{spec,corr}}$ scaling relation $\log(M_{500}/\mathrm{M_\odot}) = 1.935\log(k_\mathrm{B}T_{\mathrm{spec,corr}}$ $[\mathrm{keV}]) + 13.32$ from \simbac~\citep{houghSimbaCEvolutionThermal2024a} to estimate the $M_{500}$ of observed groups from their measured global X-ray temperatures. The calculated $M_{500}$ values are used to place each observed group into its corresponding mass bin. Due to the inherently uncertain nature of the masses calculated with this method, we permit observed groups that are sufficiently close in mass to their neighboring bins to be counted in those bins. Specifically, groups that have a value of $M_{500}$ that is up to 25\% of the relevant bin's width in logarithmic scale above or below the edges of that bin are included in the bin. This results in some groups appearing in multiple $M_{500}$ bins or being used in the calculations for more than one average profile. {All 44 \textsc{CHEERS} systems end up being included in at least one of our three mass bins, except for one very-low-mass group, NGC 5813.}

In supplementary tests, we find that setting this 25\% to 0\% results in relatively minor changes that are insignificant compared to the scatter and uncertainty already present in simulations and observations. However, we opt to keep the 25\% value, as it elevates the statistical significance from averaging.

Temperature measurements for \textsc{CHEERS} objects (including the groups from Werner et~al. \cite{wernerXMMNewtonHighresolutionSpectroscopy2006} and Grange et~al. \cite{grangeMetalContentsTwo2011}) come from de Plaa et~al. \cite{deplaaCHEERSChemicalEvolution2017}, who provide emission-weighted global temperatures adapted from Chen et~al. \cite{chenStatisticsXrayObservables2007} and Snowden et~al. \cite{snowdenCatalogGalaxyClusters2008}. For the group sample from Sarkar et~al. \cite{sarkarChemicalAbundancesOutskirts2022}, temperature values from the literature are used: Sarkar et~al. \cite{sarkarJointSuzakuChandra2021} for MKW4; Wong et~al. \cite{wongSUZAKUXRAYOBSERVATIONS2016} for Antlia; Su et~al. \cite{suENTIREVIRIALRADIUS2015} for RX J1159; and Su et~al. \cite{suSUZAKUOBSERVATIONSXRAY2013} for ESO 3060170. Ghizzardi et~al.~\citep{ghizzardiIronXCOPTracing2021a} provide hydrostatic $M_{500}$ values from Ettori et~al.~\citep{ettoriHydrostaticMassProfiles2019}. Because these masses are all well into our high-mass bin, we assume that converting their corresponding temperatures to $M_{500}$ using our $M_{500}-T_{\mathrm{spec,corr}}$ relation would not sufficiently change the masses to sit outside the high-mass bin; therefore, we do not go to the effort of finding their X-ray temperatures in the literature.

We also employ the \simbac $M_{500} - T_{\mathrm{spec,corr}}$ scaling relation to convert the $M_{500}$ bins into $T_{\mathrm{spec,corr}}$ bins (see Table~\ref{tab:num_groups}), since these values are more useful and directly applicable from an observational perspective.

\subsection{Calculating Abundance Profiles}
\label{subsec:methods_calculating_abundance_profiles}

To explore how \simbac differs from \simba in terms of the distribution of metals in the IGrM and how the results compare to observations, we investigate the averaged radial profiles of the gas-phase abundances of various key elements. Mernier et~al. \cite{mernierAstrophysicsRadialMetal2017} and Sarkar et~al. \cite{sarkarChemicalAbundancesOutskirts2022} present projected (2D) profiles, arguing that deprojection assumes spherical symmetry, an assumption that is not necessarily accurate in groups and clusters because they may exhibit non-spherical morphologies. We therefore choose to project our simulated profiles to ensure a like-to-like comparison with observations\endnote{Some studies find projecting simulated profiles to have a significant impact on their shape and/or normalization, generally decreasing and flattening the profiles, especially in the group cores (e.g.,~\citep{braspenningFLAMINGOProjectGalaxy2024,nelsonIntroducingTNGClusterSimulation2024}), whereas others find no notable differences (e.g.,~\citep{vogelsbergerUniformityTimeinvarianceIntracluster2018,altamuraEAGLElikeSimulationModels2023}). Our results follow the former findings.}. For the same reason, we use the projected profiles from Ghizzardi et~al.~\citep{ghizzardiIronXCOPTracing2021a}.

For the simulated 2D $L_X$-weighted abundance profiles of each group, all IGrM particles in a cylinder of radius $3R_{500}$ and total height $3R_{500}$ centered on each group's shrinking-sphere center are projected onto the $x$-$y$ plane\endnote{Choosing the $z$-axis as the axis of projection results in limited biasing, since any preferential alignment of the halos or spherical asymmetry should be statistically averaged out in each mass bin.}. We bin the particles by halocentric radius into 14 non-overlapping circular annuli. The edges of these bins are identically defined for every group in terms of $R_{500}$, with the first bin being in the range of $0$--$0.01R_{500}$, 10 equally log-spaced bin edges in the range of $0.01$--$1R_{500}$, and 4 in the range of $1$--$3R_{500}$. We compute X-ray luminosities and metal mass fractions for each IGrM particle, and then the emission-weighted average abundance in each annulus of each group, as described in Section~\ref{subsec:methods_xray_properties} and Equation~\eqref{eqn:ZLx}.

For the 3D mass-weighted profiles, the same procedure is carried out with no projection; instead, radial shells are spherically averaged according to Equation~\eqref{eqn:Zmass}. The same radial bin definitions are employed, with the addition of 10 more equally log-spaced bin edges in the range of $3$--$10R_{500}$.

Elements that are of special interest to the observational X-ray community include C, N, O, Ne, Mg, Si, S, Ca, Fe, and Ni~\citep{bohringerXraySpectroscopyGalaxy2010,wernerHotAtmospheresGalaxies2020,gastaldelloMetalContentHot2021}. The distributions of the abundances of these metals in galaxy groups and clusters can indicate how and when chemical enrichment occurs, what effects it will have on the future evolution of groups, including cooling and dust formation, and the efficiency of the dispersal and mixing mechanisms at play (e.g., \citep{scannapiecoFeedbackMetalEnrichment2005,tornatoreChemicalEnrichmentGalaxy2007,wiersmaChemicalEnrichmentCosmological2009,biffiHistoryChemicalEnrichment2017,biffiOriginICMEnrichment2018}, see also \citep{gastaldelloMetalContentHot2021,mernierChemicalEnrichmentGroups2022} and the references therein). We note that some of these metals, such as C, N, Ne, and Ni, are more difficult to measure due to {the lack of strong, well-separated spectral lines in the observed X-ray bands}, resulting in limited observations.

\subsubsection{Stacking Profiles}
\label{subsubsec:methods_stacking_profiles}

To place any sort of statistically significant constraints or draw qualitative or quantitative conclusions, we must make use of average profiles. Simulations and observations suffer from opposite limitations in two major ways in the context of this study, making comparisons slightly more challenging.

First, given the shape of the halo mass function, a fixed volume contains more lower- than higher-mass galaxy groups. Observational samples tend to be weighted towards---or at least have greater statistics for---massive groups and clusters because their X-ray surface brightnesses are larger. Due to this bias, catalogs are more complete at larger halo masses (e.g.,~\citep{mernierAstrophysicsRadialMetal2017,ghizzardiIronXCOPTracing2021a}). On the other hand, the statistics of cosmological simulations are limited by the volume of the simulation box. Thus, simulation samples have greater statistical significance for lower-mass groups than higher-mass groups, opposite to observational~samples.

\textls[-10]{Second, the outskirts of the observed groups are relatively X-ray dim compared to their cores, where the gas tends to be more dense (e.g.,~\citep{mulchaeyXRayPropertiesGroups2000}). Therefore, uncertainties in and the scatter of inferred abundances are often greater in the group outskirts than in the cores. In simulations, however, one of the main factors impacting scatter is the number of resolution elements present in a given annulus or shell: the fewer gas particles, the worse the number statistics, resulting in a greater Poisson uncertainty (see \citep{powerInnerStructureLCDM2003,ludlowNumericalConvergenceSimulations2019} for numerical convergence studies). Since group cores are far more compact than the extended outskirts, there are fewer gas particles in the inner regions. This results in the profiles having larger uncertainties and scatter in the core than the rest of the group, contrary to what is found for observed profiles}.

For these reasons, we use different radial bins and bin sizes for simulations compared to observations. By necessity, for decent statistics, the inner radial bins in simulations are comparatively larger in a log-scaled sense than the outer ones (see Section~\ref{subsec:methods_calculating_abundance_profiles}), and vice versa for the observational profiles. We also take slightly different approaches to determining averages and representational uncertainties in simulated versus observed group samples, as described in the following two sections.

\subsubsection{Stacking Observations}
\label{subsubsec:methods_stacking_obs_profiles}

\textls[-10]{{To create average observed abundance profiles in each $M_{500}$ bin, we adapt the procedure described by Mernier et~al. \cite{mernierAstrophysicsRadialMetal2017}, first proposed by Leccardi and Molendi \cite{leccardiRadialMetallicityProfiles2008}. In the study of Mernier et~al. \cite{mernierAstrophysicsRadialMetal2017}, since the spectral analysis to compute the abundance profiles of individual groups was performed within annuli of fixed angular sizes, regardless of the distance or redshift, the resulting profiles rescaled to each group's $R_{500}$ have different radial bins. Thus, this method requires choosing a set of ``reference'' radial bins in units of $R_{500}$ for the final average abundance profiles. We employ the radial bins from Mernier et~al.'s ``group'' subsample \cite{mernierAstrophysicsRadialMetal2017}, with bin edges [0, 0.009, 0.024, 0.042, 0.064, 0.1, 0.15, 0.26, 0.97]$R_{500}$.}}

In each $M_{500}$ bin and reference radial bin, we then compute the median abundances weighted by the individual observation's statistical errors from spectral fitting, as well as a weighting factor between 0 and 1 representing the linear overlapping geometric area fraction of the $i${th} reference radial bin on the $j${{th}} annulus belonging to the $k${th} galaxy group\endnote{This overlap factor $w$ is calculated as follows: if the radial bin is fully contained in the reference radial bin (inclusive), $w=1$; if the bin is fully outside the reference bin, $w=0$; otherwise, if the bin and reference bin partly overlap, $w$ is the fraction of the radial bin overlapping with the reference bin.}. {The scatter of the average profile is given by the $16${th} and $84${th} percentiles weighted by the same quantities, as these converge to $1\sigma$ limits in the case of a Gaussian distribution. We find no qualitative differences between the $16${th} and $84${th} percentiles compared to the quartiles ($25${th} and $75${th} percentiles) in terms of the agreement/discrepancy across simulations and observations; quantitatively, there is a minor shrinking of the error bars.}

This method reduces the effect of uncertain measurements on the final average profile, and increases the weight of measurements with excellent data quality (e.g., from a very bright group, or from an instrument with higher resolution or sensitivity). As recommended by Mernier et~al. \cite{mernierAstrophysicsRadialMetal2017}, we include radial bins with unphysical negative abundances derived from spectral fitting so as not to bias the averages, but discard certain radial bins from specific groups that do not meet their standards, due to bad data quality, contamination (e.g., from AGN), or poor spectral fitting (details can be found in Mernier et~al.~\citep{mernierAstrophysicsRadialMetal2017}).

In the study of Mernier et~al. \cite{mernierAstrophysicsRadialMetal2017}, Fe was measured in simultaneous \emph{XMM-Newton}/EPIC MOS+pn ``global'' spectral fits, while O could only be measured with the MOS instruments. The rest of the elements were measured separately with both MOS and pn (``local'' fits). While Mernier et~al. \cite{mernierAstrophysicsRadialMetal2017} devised a method to combine the measurements from the MOS and pn instruments, we opt to keep these separate for the elements that have both, since they are sometimes found to provide systematically different values. As the true abundances are not known a priori, there is no way to pick one of these instruments over the other. Therefore, we take any spread between the MOS and pn stacked profiles, on top of their individual scatter, to be representative of the uncertainty.

\textls[-15]{We employ the same method of error-weighted averaging for the core abundance results from Mao et~al. \cite{maoNitrogenAbundanceXray2019} and Fukushima et~al. \cite{fukushimaNeMgFe2023}, albeit with only a single reference radial bin at small radii. All groups in the X-COP sample of Ghizzardi et~al.~\citep{ghizzardiIronXCOPTracing2021a} fall into our high-mass bin; therefore, we simply use the mean profiles reported directly in their paper\endnote{Ghizzardi et~al.~\citep{ghizzardiIronXCOPTracing2021a} find negligible differences between the mean and median Fe profiles of the X-COP sample.}. We do not stack the profiles from Sarkar et~al. \cite{sarkarChemicalAbundancesOutskirts2022}, as their sample only consists of four objects}.

In supplementary tests, we find minimal changes in the averaged observed abundance profiles and little impact on the agreement between simulated and observed abundance profiles when altering some of the choices we make for averaging. This includes the choice of reference radial bins, whether negative abundances are excluded from averaging, and whether certain groups with excellent data quality (that may bias the average profile towards their values) are excluded.

\subsubsection{Stacking Simulations}
\label{subsubsec:methods_stacking_sim_profiles}

Ideally, for a consistent comparison, simulated profiles would be averaged in the same manner as observations. However, the uncertainties in the simulated abundance profiles decrease with increasing halo mass, as the number of particles increases\endnote{Because the number of groups decreases with increasing $M_{500}$ in cosmological simulations, the scatter in the median profiles also increases; however, this does not necessarily introduce any bias.}. Hence, performing an error-weighted average would bias the resulting abundance profiles towards the higher-mass groups. Thus, we do not weight the average abundance profiles from simulations by uncertainty. Rather, we calculate the abundance profiles of each group individually, and then find the median values of each radial bin in each $M_{500}$ bin. We again take the $16${th} and $84${th} percentiles to represent the uncertainty.

To address numerical resolution concerns, particularly in the high-density/low-particle-number group cores, we implement four cuts, described in the following list. According to Dav\'e et~al. \cite{daveMUFASAGalaxyFormation2016}, \simba and \simbac employ \textsc{GIZMO} with a 64-neighbor cubic spline kernel for hydrodynamics, and, similarly, a 64-particle kernel for adaptive gravitational softening. For this reason, here, ``resolved'' refers to a radial bin that contains at least 64 resolution elements.

\begin{enumerate}
    \item Unresolved individual radial bins of individual groups are excluded from averaging. In the case of projected profiles, if a 2D annulus has any contributions from any unresolved 3D shells in the group, the annulus is also considered unresolved. The reason for this is that, spectroscopic observations being effectively X-ray emission weighted and $L_X$ increasing towards the center of a group, the abundances in annuli have greater contributions from the inner spherical shells. Therefore, in our simulations, if these inner shells are unresolved, it can impact the resultant projected abundance values significantly\endnote{We find that using the total number of particles projected into an annulus as the resolution cut criterion for each annulus results in insignificant changes to the projected abundance profiles in comparison to using the number of particles in each spherical shell that contributes to the annulus (our approach discussed in this text). The former method generally leads to slightly higher median amplitudes in the innermost radial bins, and---as expected---produces far fewer unresolved radial bins (i.e., the profiles extend further in towards the group cores). Any qualitative conclusions we draw are robust to the choice of resolution cut method.}.
    \item Per group, radial bins that are thinner than the \emph{{physical}} minimum gravitational softening length are excluded. The physical value is the Plummer-equivalent value multiplied by 2.8, to account for the conversion to the actual smoothing kernel radius~\citep{daveMUFASAGalaxyFormation2016}. Here, this gives $0.5\,\text{h}^{-1}$ {kpc ``Plummer-equivalent''} = $0.5/0.68 \times 2.8 \, \mathrm{kpc} \approx 2 \, \mathrm{kpc}$.
    \item Similar to Braspenning et~al. \cite{braspenningFLAMINGOProjectGalaxy2024}, we require at least 16\% of the groups in a radial bin to be resolved. Below this point, the $1\sigma$ error can no longer be reliably determined. If this condition is not met, the radial bin is fully ignored and does not appear in the averaged profile. Note that this may result in the averaged profiles for different mass bins, and for \simba and \simbac, having different inner stopping radii.
    \item There must be a minimum of five resolved individual radial bins contributing to a given radial bin of the average profile; otherwise, the radial bin is ignored overall. This condition is imposed to ensure that radial bins that undergo the prior cuts still contain sufficient statistics to consider the average as representative of the population.
\end{enumerate}

\section{Results}
\label{sec:results}

Our results focus on IGrM abundance profiles. We compare \simbac and \simba to each other and to the observational samples outlined in Section~\ref{subsec:methods_obs_samples}. Kobayashi et~al. \cite{kobayashiOriginElementsCarbon2020} describe the setup of the \texttt{Chem5} model and the proportion of each element that originates from different astrophysical sources across cosmic time, which, for the purposes of the present study, constitute AGB stars, core-collapse SNe (SNe II and HNe), and SNe Ia. Based on their results, and focusing only on those metals observable in X-ray and for which we have observational data, we split up the metals according to their origins as follows:

\begin{itemize}
    \item The iron-peak elements \textbf{Fe and Ni} are produced largely by SNe Ia. In \texttt{Chem5}, about 60\% of Fe comes from SNe Ia, with the rest mainly produced by HNe. Also, an appreciable amount of Ni comes from core-collapse SNe. We focus heavily on Fe, since it is the most robustly constrained metal from X-ray observations, and has been shown to be a good tracer of overall IGrM metallicity~\citep{renziniIronTracerGalaxy1997,urbanUniformMetallicityOutskirts2017,vogelsbergerUniformityTimeinvarianceIntracluster2018,pearceRedshiftEvolutionHot2021,gastaldelloMetalContentHot2021,mernierChemicalEnrichmentGroups2022}.
    \item \textbf{C and N} are produced primarily by AGB stars, with considerable amounts from core-collapse SNe (up to $\sim$$40$\% for C at $z \sim 0$).
    \item The light $\alpha$ elements \textbf{O, Ne, and Mg} originate largely from core-collapse SNe.
    \item The heavy $\alpha$ elements \textbf{Si, S, and Ca}, similar to the other $\alpha$ elements, are mainly produced by core-collapse SNe, but with a significant proportion ($\sim$20{--}40\%) originating from SNe Ia.
\end{itemize}

For each grouping of metals, we compare the \simba abundance profiles to those of observations, followed by a comparison of \simbac and observations (emphasizing the differences found for \simbac), and finally compare the \simbac and \simba abundance~profiles.

\subsection{Iron-Peak Elements: Fe and Ni}
\label{subsec:fe_profiles}

\begin{figure}[H]
    \begin{adjustwidth}{-\extralength}{0cm}
    \centering
    \includegraphics[width=0.99\linewidth]{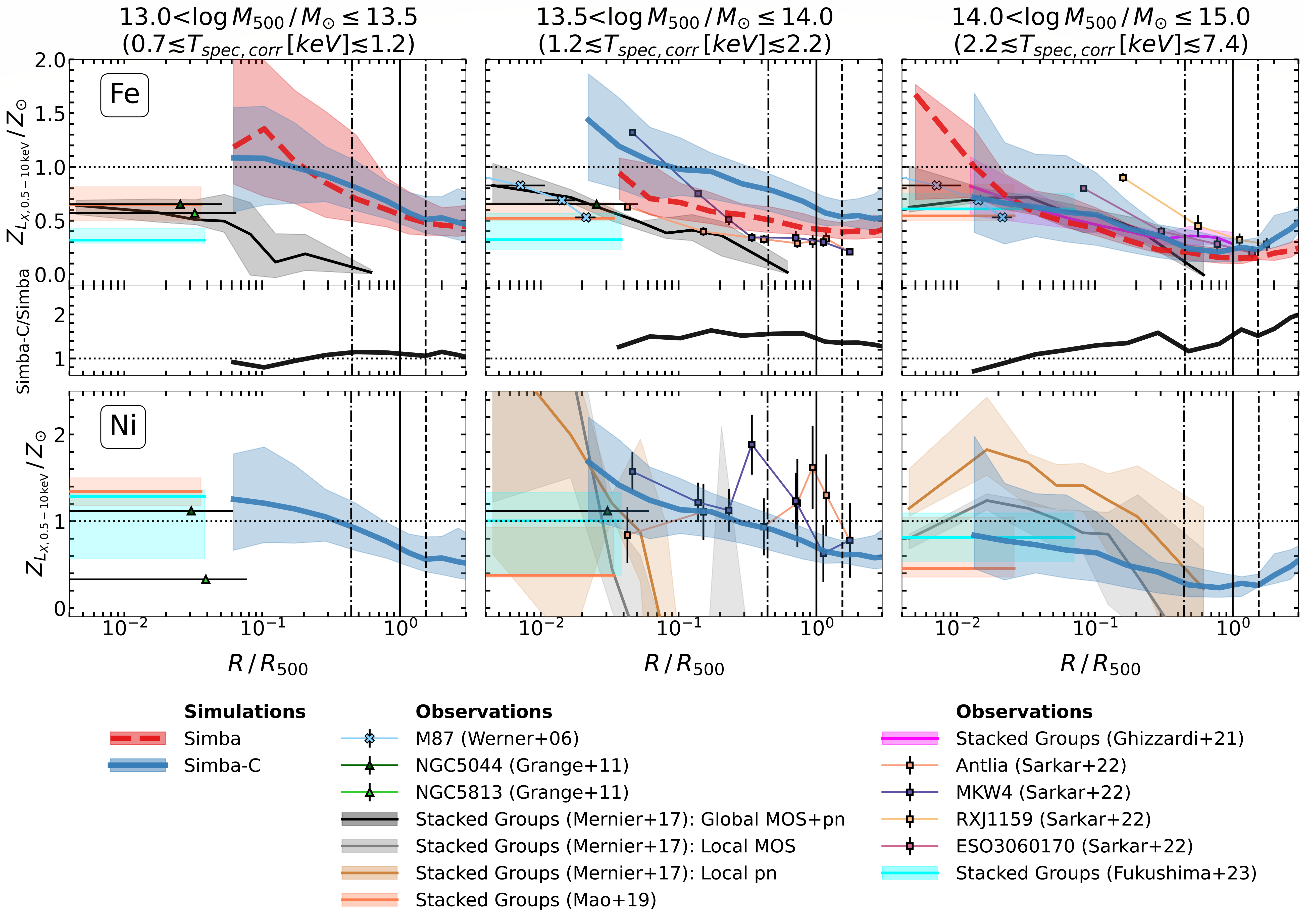}
    \end{adjustwidth}
    \caption{Median 2D projected Fe (\emph{{upper row}}) and Ni (\emph{{lower row}}) abundance profiles from \simba and \simbac, as well as a compilation of observational results. Note that \simba does not track Ni. Abundance profiles of individual groups are emission weighted by the IGrM X-ray luminosity in the 0.5--10 keV band, and are scaled to the solar abundances of Asplund et~al. \cite{asplundChemicalCompositionSun2009}. Shaded regions represent the 16th--84{th} percentiles in the scatter. The three columns split all data into our three bins in group mass $M_{500}$, with their corresponding approximate temperatures $T_{\mathrm{spec,corr}}$ indicated. The vertical solid line is $R_{500}$, and the vertical dot-dashed and long-dashed lines are our approximations for $R_{2500}$ and $R_{200}$, respectively, using the median values from our two simulated samples. The horizontal dotted line is solar abundance $Z_{\odot}$. The radial axes extend in the range of $0.004$--$3R_{500}$. The upper limit, just above $R_{200}$, cuts out the IGM, as we focus primarily on the IGrM in this study. {The lower $40\%$ of each panel in the top row shows the ratio of the \simbac to \simba abundance profiles, highlighting their relative amplitudes across radii.}}
    \label{fig:fe_ni_profiles}
\end{figure}

\simbac tracks all of the Fe-peak elements, while \simba only tracks Fe. To the best of our knowledge, on top of Fe, X-ray observational data of Fe-peak elemental abundances in the IGrM are only available for Cr, Mn, and Ni\endnote{For Cr and Mn, we can only find abundance \emph{{ratios}} (e.g., X/Fe) in the literature, not \emph{absolute} abundances. However, the simulated Cr and Mn abundance profiles have similar amplitudes, shapes, and trends with $M_{500}$ as Ni.}. Figure~\ref{fig:fe_ni_profiles} shows the simulated average IGrM Fe and Ni abundance profiles of all selected galaxy groups in \simba and \simbac, as well as the observational samples. The lower $40\%$ of each individual panel plots the ratio of the \simbac to \simba abundance profiles, which is useful for a direct comparison of the amplitude and shape of the two simulated profiles.

Figure~\ref{fig:fe_profiles-high_mass_bin-simba+obs-simbac+obs} shows the simulated and observed Fe abundance profiles in the high-mass bin, separated by simulation (see caption), since this panel in Figure~\ref{fig:fe_ni_profiles} is quite crowded.

\subsubsection{\simba--Observation Comparison}
\label{subsubsec:results_fe_ni_simba_obs_comparison}

On average, the low-mass \simba groups (leftmost bin) are discrepant with observations at $R \gtrsim 0.1R_{500}$, being around a factor of 2{--}5 too high. \simba and observations become more similar in amplitude in the intermediate-mass bin (middle column) through a slight decrease and increase, respectively. With respect to \textsc{CHEERS}~\citep{mernierAstrophysicsRadialMetal2017}, however, \simba is still inconsistent at $>$1$\sigma$.

In the high-mass bin (rightmost column), in the outskirts at $R\lesssim R_{200}$, \simba and observations converge to $\sim$0.1{--}0.3$ \, Z_{\mathrm{Fe},\odot}$, similar to other studies of massive groups and clusters in the literature, both simulation based (e.g.,~\citep{barnesClusterEAGLEProjectGlobal2017,biffiHistoryChemicalEnrichment2017,biffiOriginICMEnrichment2018,vogelsbergerUniformityTimeinvarianceIntracluster2018,braspenningFLAMINGOProjectGalaxy2024}) and observational (e.g.,~\citep{urbanUniformMetallicityOutskirts2017,lovisariNonuniformityGalaxyCluster2019,ghizzardiIronXCOPTracing2021a,sarkarChemicalAbundancesOutskirts2022,gatuzzChemicalEnrichmentICM2023}). This can be seen clearly in Figure~\ref{fig:fe_profiles-high_mass_bin-simba+obs-simbac+obs}. Notably, the median \simba Fe abundance rises steeply towards the center at $R \lesssim 0.03R_{500}$, whereas observations remain flatter and sub-solar, although the lower bound of the uncertainty limits still overlaps with~observations.

\subsubsection{\simbac--Observation Comparison}
\label{subsubsec:results_fe_ni_simbac_obs_comparison}

In the low-mass bin, the \simbac Fe profile is similar to that of \simba, and equally inconsistent with observations. Unlike \simba, however, \simbac changes very little from the low- to intermediate-mass bins, in which it has a higher-amplitude median profile than \simba and is in less agreement with observations.

\begin{figure}[H]
    \begin{adjustwidth}{-\extralength}{0cm}
    \centering
    \includegraphics[width=0.99\linewidth]{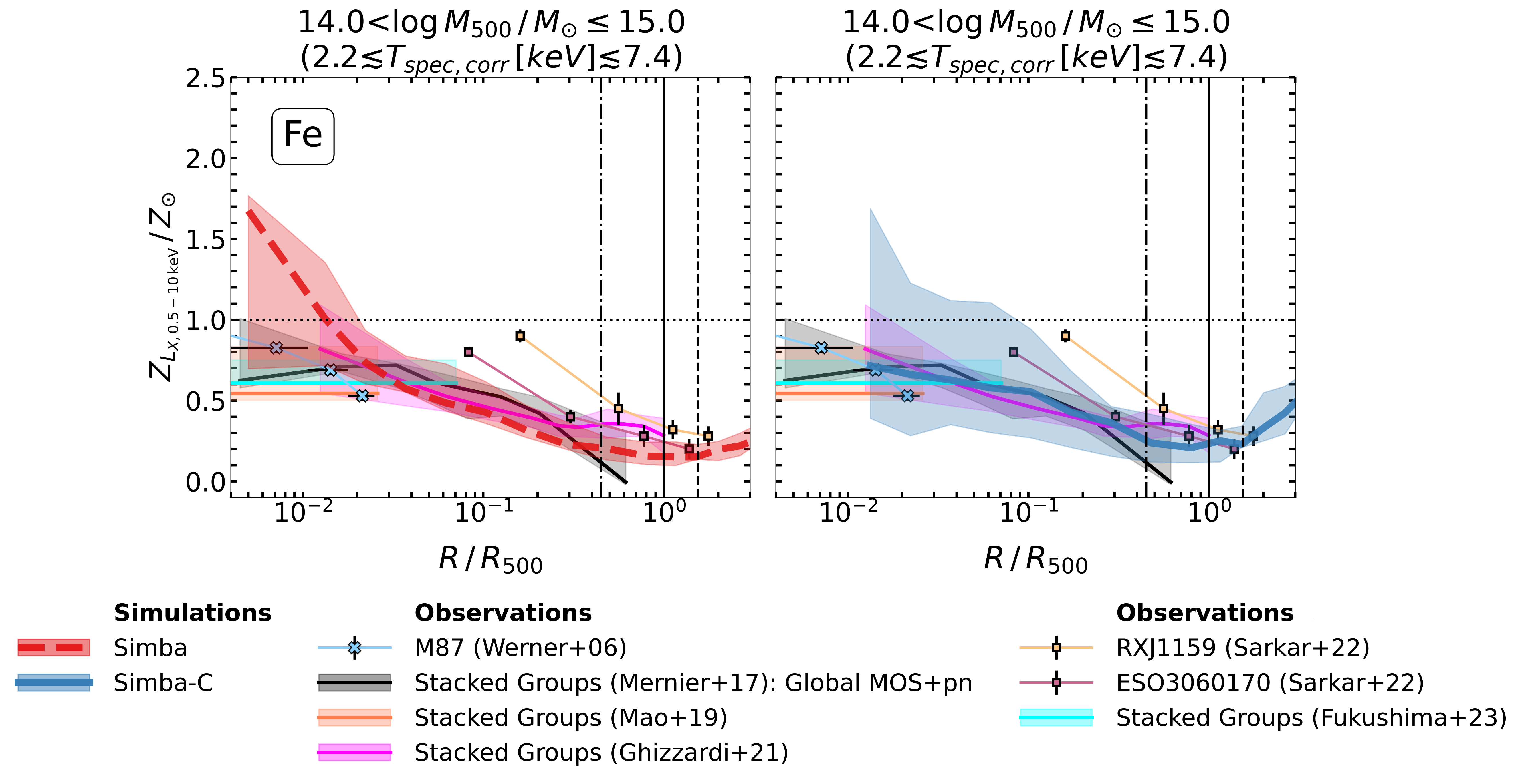}
    \end{adjustwidth}
    \caption{Projected $L_X$-weighted Fe abundance profiles of \simba and \simbac in the high-mass bin, separated by simulation: \simba (\emph{{left}}) and \simbac (\emph{{right}}). Both subplots show the same set of observational samples. This figure is meant to separate out and clarify the comparison of the simulated and observed abundance profiles from the first row, third column in Figure~\ref{fig:fe_ni_profiles}, which is quite crowded. All other characteristics of the plot are the same as in Figure~\ref{fig:fe_ni_profiles}. The vertical solid line is $R_{500}$, and the vertical dot-dashed and long-dashed lines are our approximations for $R_{2500}$ and $R_{200}$, respectively, using the median values from our two simulated samples. The horizontal dotted line is the solar abundance $Z_{\odot}$.}
    \label{fig:fe_profiles-high_mass_bin-simba+obs-simbac+obs}
\end{figure}

\textls[-15]{In high-mass groups, \simbac exhibits excellent agreement with observations across all radii\endnote{Note that the innermost radial bin of each group in the Sarkar et~al. \cite{sarkarChemicalAbundancesOutskirts2022} sample includes all data from the center of the group out to the first bin edge, which tend to be a fair bit larger than the first bin edges of the simulated and other observed profiles. In this way, there are contributions to the inner Sarkar et~al. \cite{sarkarChemicalAbundancesOutskirts2022} radial bins from the core, where the abundances tend to be higher, but, due to the log scaling of the radial axes, these values then appear at larger radii.}. As with \simba, \simbac converges with observations in the outskirts. Unlike \simba, \simbac remains flatter, sub-solar, and in agreement with observations at \mbox{$R \lesssim 0.03R_{500}$} (see Figure~\ref{fig:fe_profiles-high_mass_bin-simba+obs-simbac+obs}). At this mass scale, \simbac is a clear improvement over \simba}.

The \simbac profiles for Ni follow similar trends to Fe (i.e., abundances at $R \gtrsim 0.1R_{500}$ decreasing and core abundances flattening from the intermediate- to high-mass bin), and the two have similar amplitudes. This is expected because Ni is itself an Fe-peak element, and the ratio of the Fe and Ni yields is roughly constant across stellar metallicity for both core-collapse SNe and SNe Ia. Observational profiles for Ni, especially for lower-mass groups, are highly uncertain {because the Ni-L shell emission lines are largely unresolved with current CCD instruments (e.g.,~\citep{guXRaySpectroscopyGalaxy2018,gastaldelloMetalContentHot2021})}\endnote{\textsc{CHEERS} has measurements of Ni abundance profiles at low group masses, but, as they are quite limited, we have opted not to show them.}. Although the \simbac abundance profiles of Fe and Ni are similar in amplitude, by extrapolating radially inwards, the Ni profiles better agree with observations in the low-mass bin than those of Fe. This is due to a greater abundance of Ni compared to Fe in the cores of the observed lower-mass groups. In the intermediate-mass bin, \simbac appears to produce slightly high core abundances (consistent with the more uncertain results of Mernier et~al.~\citep{mernierAstrophysicsRadialMetal2017}), but agrees well with the outskirt values, and is broadly consistent with observations in the high-mass bin. The agreement in the intermediate-mass bin at $R \gtrsim 0.1R_{500}$ should be interpreted as preliminary due to low observational statistics; more high-quality observations are required.

\vspace{-4pt}
\vspace{-3pt}

\subsubsection{\simbac--\simba Comparison}
\label{subsubsec:results_fe_ni_simbac_simba_comparison}

At $R \gtrsim 0.1R_{500}$, \simba Fe abundance gradually decreases with increasing $M_{500}$. Simultaneously, the \simba abundance profiles become more centrally peaked, potentially indicating that the deepening gravitational potentials are more efficiently retaining metals in the inner regions.

As stated, \simbac, in contrast, alters minimally from the low- to intermediate-mass bins, such that the two simulations become largely discrepant at $\gtrsim$$1\sigma$. This is followed by an abrupt decrease in the \simbac abundance profile amplitude in the high-mass bin, with the profile amplitude and shape being broadly comparable to that of \simba for $R \gtrsim 0.1R_{500}$ but significantly flatter at $R \lesssim 0.1R_{500}$\endnote{Since there are only nine groups in each simulated sample in the high-mass bin, we test removing different groups from this bin, and find that the median profiles in \simba and \simbac change minimally, preserving the trends we find and confirming that they are statistically robust.}$^\mathrm{,}$\endnote{In the high-mass bin, due to the resolution cuts, \simbac ``stops'' one radial bin above \simba. Without these cuts, the trends already found are further exaggerated, indicating that it is unlikely that \simbac would show a rapid increase in core Fe abundance (similar to \simba) if the simulation resolution were enhanced.}.

Additionally, at high masses, the outer regions of \simbac groups contain, on average, somewhat greater abundances of Fe than in \simba. Combined with the different core abundances between \simba and \simbac, this suggests that, in the high-mass groups of \simbac, Fe is expelled to larger radii more efficiently, whereas, in \simba, the Fe is confined substantially more to the core region. On the other hand, in low-mass groups, outflows appear to be shuttling metals to the group outskirts more equally in the two simulations.

Overall, the improved agreement for \simbac at high masses notwithstanding, the simulation--observation discrepancy in Fe abundance profiles at low masses for both \simba and \simbac indicates that there exist persistent issues in our simulations at the group scale that still need to be addressed.

\subsection{C and N}
\label{subsec:c_n_profiles}

\begin{figure}[H]
    \begin{adjustwidth}{-\extralength}{0cm}
    \centering
    \includegraphics[width=0.99\linewidth]{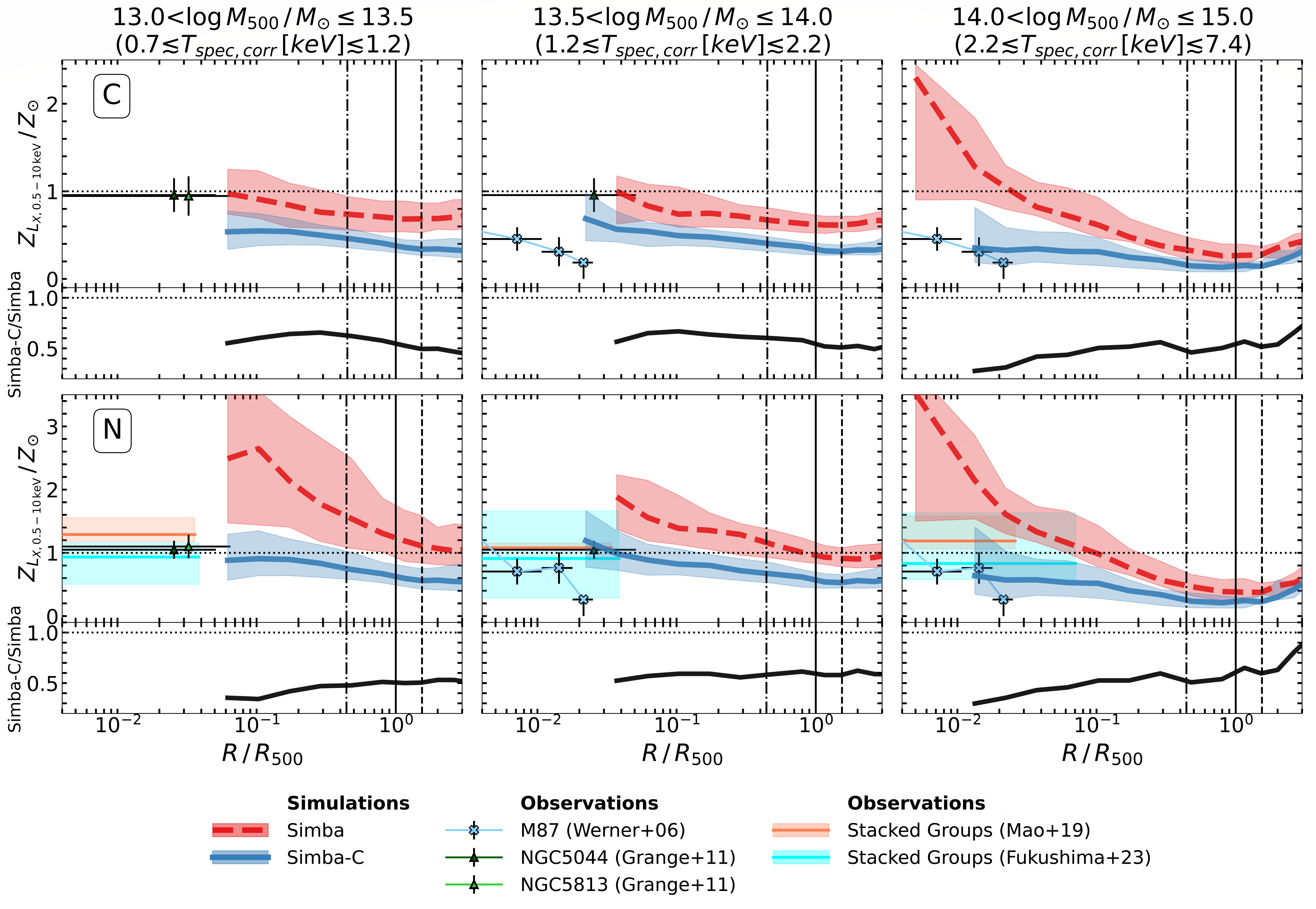}
    \end{adjustwidth}
    \caption{Same as Figure~\ref{fig:fe_ni_profiles} but for C (\emph{{upper row}}) and N (\emph{{lower row}}). The vertical solid line is $R_{500}$, and the vertical dot-dashed and long-dashed lines are our approximations for $R_{2500}$ and $R_{200}$, respectively, using the median values from our two simulated samples. The horizontal dotted line is solar abundance $Z_{\odot}$.}
    \label{fig:c_n_profiles}
\end{figure}

Figure~\ref{fig:c_n_profiles} depicts the C and N abundance profiles in \simba and \simbac, and core abundances of these metals in observations.

\vspace{-3pt}
\subsubsection{\simba--Observation Comparison}
\label{subsubsec:results_c_n_simba_obs_comparison}

\simba produces group cores over-enriched in N by up to factors of $\sim 4$ in the low- and high-mass bins. The disagreement is less extreme in the intermediate bin but still evident. {We note, however, that observations also find highly super-solar abundances of N in the IGrM. For example, some groups in the Mao et~al. \cite{maoNitrogenAbundanceXray2019} and Fukushima et~al. \cite{fukushimaNeMgFe2023} samples have N abundances of up to $\sim$2$Z_\mathrm{N,\odot}$, while NGC 1404 was found to have abundances of around $2-3Z_\mathrm{N,\odot}$ \cite{mernierCycleMetalsInfalling2022}.}

Observations of IGrM C abundances are generally too limited to draw conclusions confidently. At lower masses, extrapolating radially inwards, \simba appears to match observations, while, similar to N, it generally over-enriches high-mass groups in terms of C. A more comprehensive study, collecting all data from the literature and/or using the upcoming observations from \emph{XRISM}, will be carried out in the future.

\subsubsection{\simbac--Observation Comparison}
\label{subsubsec:results_c_n_simbac_obs_comparison}

Unlike \simba, \simbac agrees well with observations of N core abundances across all masses (extrapolating radially inward in the low-mass bin), and with C core abundances in the intermediate- and high-mass bins. However, \simbac appears to under-enrich low-mass groups with C, making it slightly less consistent with observations than \simba.

C and N are primarily produced by AGB stars. Therefore, the overhauled treatment of AGB stars in \textsc{Chem5} in \simbac improves concretely on the enrichment of the IGrM with N, and potentially with C, in particular in group cores.

\subsubsection{\simbac--\simba Comparison}
\label{subsubsec:results_c_n_simbac_simba_comparison}

Figure~\ref{fig:c_n_profiles} shows that both C and N feature one similar trend to Fe: from the two lower-mass bins to the high-mass bin, the \simba profiles become highly centrally concentrated, whereas the \simbac profiles remain flat in the core.

In contrast to Fe though, at all $M_{500}$ and all radii, the profiles of \simba have statistically ($\gtrsim$$1\sigma$) larger amplitudes than those of \simbac. This is particularly clear in the inner regions, where \simba largely produces inwardly increasing abundance profiles while the \simbac profiles are flatter and may even hint at a ``turn over'' moving into the core.

\subsection{Light $\alpha$ Elements: O, Ne, and Mg}
\label{subsec:o_ne_mg_profiles}

\begin{figure}[H]
    \begin{adjustwidth}{-\extralength}{0cm}
    \centering
    \includegraphics[width=0.99\linewidth]{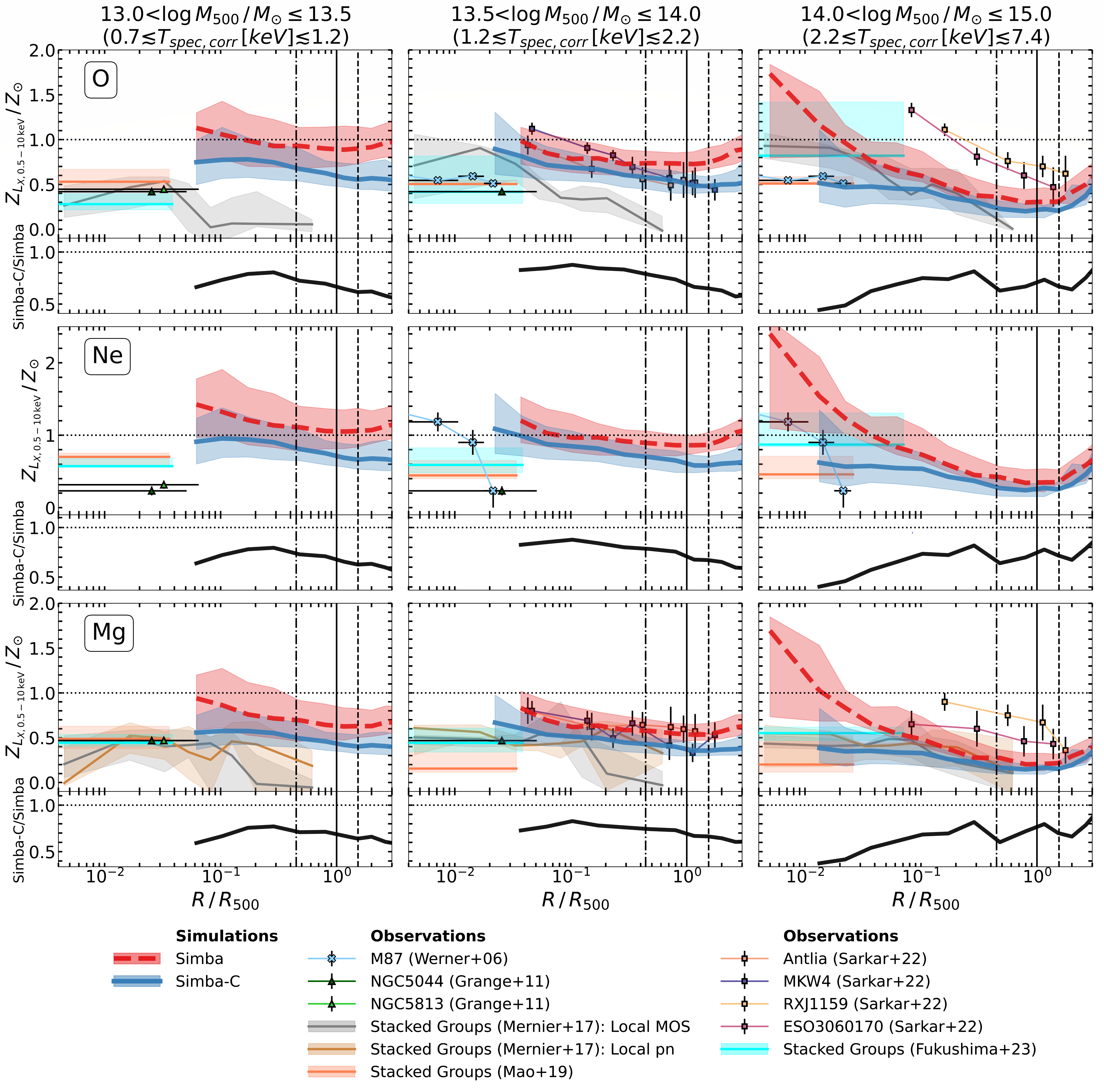}
    \end{adjustwidth}
    \caption{Same as Figure~\ref{fig:fe_ni_profiles} but for O (\emph{{upper row}}), Ne (\emph{{middle row}}), and Mg (\emph{{lower row}}). The vertical solid line is $R_{500}$, and the vertical dot-dashed and long-dashed lines are our approximations for $R_{2500}$ and $R_{200}$, respectively, using the median values from our two simulated samples. The horizontal dotted line is solar abundance $Z_{\odot}$.}
    \label{fig:o_ne_mg_profiles}
\end{figure}

Figure~\ref{fig:o_ne_mg_profiles} shows the IGrM abundance profiles of O, Ne, and Mg for \simba, \simbac, and observations.

\subsubsection{\simba--Observation Comparison}
\label{subsubsec:results_o_ne_mg_simba_obs_comparison}

In the low-mass bin, for all three metals, \simba produces abundance profiles that are too high in amplitude compared to observations. Additionally, the \simba profiles show an inwardly increasing trend toward super-solar abundances, taking it further away from the sub-solar abundances observed in galaxy group cores.

While the same is true in the intermediate-mass bin, the profiles from \simba and the data from \emph{XMM-Newton} are slightly closer. For O and Mg, the observations from the study of Sarkar et~al. \cite{sarkarChemicalAbundancesOutskirts2022} with \emph{Suzaku} are relatively consistent with \simba, bounding the simulation from above (at least at $R \lesssim R_{2500}$ for O). On the other hand, the \simba profiles exhibit an upturn in abundance at $R \gtrsim R_{500}$, a feature that again brings them away from the radially flat or decreasing trends of outskirt observations of O and Mg abundances.

In the high-mass bin, \simba generally produces overly enriched cores at $R \lesssim 0.1R_{500}$, but is still more or less consistent with observations. It is again bounded by the two observational samples at $R \gtrsim 0.1R_{500}$, the CHEERS sample~\citep{mernierAstrophysicsRadialMetal2017} below and the Sarkar et~al.~\cite{sarkarChemicalAbundancesOutskirts2022} sample above, respectively.

\subsubsection{\simbac--Observation Comparison}
\label{subsubsec:results_o_ne_mg_simbac_obs_comparison}

At low group masses, \simbac shows hints of a central downturn in abundance, which is non-existent in \simba. Extrapolating this trend inward suggests that \simbac may agree with observations of core abundances where \simba does not. Although both simulations produce profiles that are too high in O abundance at $R \gtrsim 0.1R_{500}$ compared to stacked group observations~\citep{mernierAstrophysicsRadialMetal2017}, \simbac's lower amplitude is regardless an improvement over \simba. Furthermore, for the same reason, while \simba sits largely discrepant with observations of Mg at $R \gtrsim 0.1R_{500}$, \simbac shows considerably more overlap.

In the intermediate-mass bin, the central downturn in \simbac has disappeared, but \simbac remains more consistent than \simba due to its lower abundance profile amplitudes. In particular, the \simbac outskirt O and Mg abundances at $R \gtrsim 0.1R_{500}$ are within the spread of observations more consistently than \simba and over a larger range of radii. Further, \simbac does not show the upturn in abundances at large radii found for \simba.

High-mass \simbac groups are again in excellent agreement with observations across the range of radii investigated. Note that observed central O abundances have more diversity than the other $\alpha$ elements, permitting \simba to also be consistent with these observations, although the shallower slope of \simbac remains a better match.

\vspace{-3pt}

Overall, \simbac is in better agreement than \simba with observations of O, Ne, and Mg abundances at all group masses.

\subsubsection{\simbac--\simba Comparison}
\label{subsubsec:results_o_ne_mg_simbac_simba_comparison}

As with C and N, \simbac has lower amplitude and generally flatter profiles than \simba. \simba therefore constrains its IGrM metals to a greater extent in both low-mass and massive group cores, while \simbac distributes them more uniformly. This is unsurprising because a significant fraction of C and N is produced by core-collapse SNe, the main producer of O, Ne, and Mg, so the stellar winds and SNe feedback can spread these metals similarly throughout the IGrM.

\subsection{Heavy $\alpha$ Elements: Si, S, and Ca}
\label{subsec:si_s_ca_profiles}

\begin{figure}[H]
    \begin{adjustwidth}{-\extralength}{0cm}
    \centering
    \includegraphics[width=0.99\linewidth]{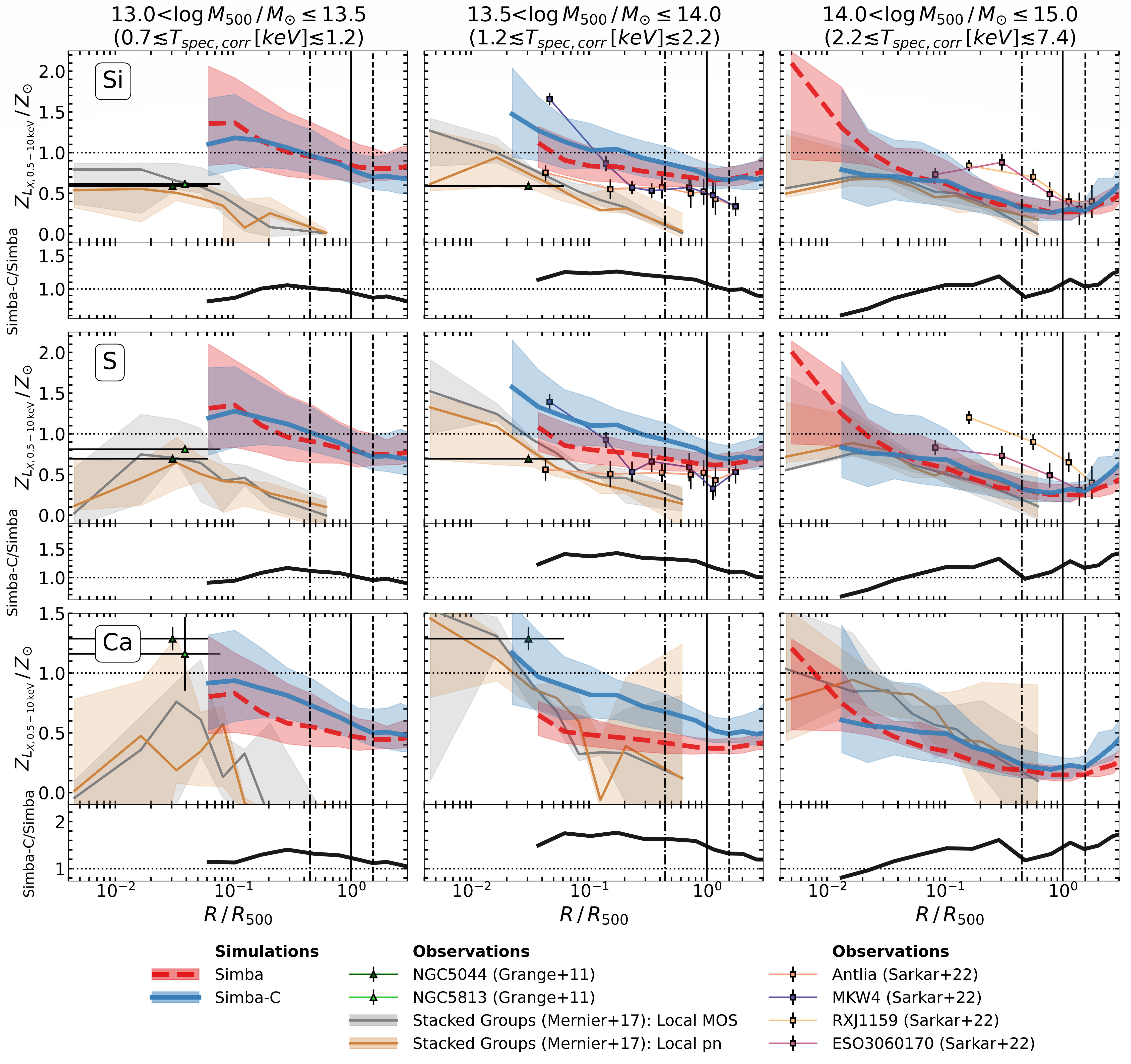}
    \end{adjustwidth}
    \caption{Same as Figure~\ref{fig:fe_ni_profiles} but for Si (\emph{{upper row}}), S (\emph{{middle row}}), and Ca (\emph{{lower row}}). The vertical solid line is $R_{500}$, and the vertical dot-dashed and long-dashed lines are our approximations for $R_{2500}$ and $R_{200}$, respectively, using the median values from our two simulated samples. The horizontal dotted line is solar abundance $Z_{\odot}$.}
    \label{fig:si_s_ca_profiles}
\end{figure}

Figure~\ref{fig:si_s_ca_profiles} presents the simulated and observed abundance profiles of Si, S, and Ca\endnote{Observations of Ar abundance in the IGrM exist; however, \simba does not track Ar, and the Ar profiles exhibit similar amplitudes, shapes, and trends with $M_{500}$ as Si, S, and Ca.}.

\vspace{-46pt}

\subsubsection{\simba--Observation Comparison}
\label{subsubsec:results_si_s_ca_simba_obs_comparison}

The Si and S profiles show a similar behavior to Fe. {\simba produces abundance profiles in the low-mass bin that have a high amplitude and are discrepant with observations. The \simba abundance profile amplitudes drop slightly from the low- to the intermediate-mass bin, resulting in \simba's increased agreement with observations. The abundance profiles in the high-mass groups of \simba are steep in the cores, and thus largely do not agree with observations, but are quite consistent with observations at $R \gtrsim 0.1R_{500}$.} This overarching similarity is understandable since, like Fe, a modest fraction of the production of Si and S is from SNe Ia according to Kobayashi et~al. \cite{kobayashiOriginElementsCarbon2020}.

Compared to Si and S, the simulated Ca abundance profiles are reduced in amplitude relative to the observed profiles. The effect is an overall improved agreement of \simba and observations in the two lower-mass bins, and a trade-off in the high-mass bin. Specifically, in the intermediate-mass bin, \simba agrees well with observations at $R \gtrsim 0.03R_{500}$. In the high-mass bin, while the shape of the \simba Ca abundance profile does not match that of observations, and its amplitude outside the core largely falls somewhat below observed abundances, the simulated and observed core abundances overlap.

\subsubsection{\simbac--Observation Comparison}
\label{subsubsec:results_si_s_ca_simbac_obs_comparison}

Like \simba, the \simbac Si and S abundance profiles follow the same trends as the \simbac Fe profiles. This includes substantial discrepancy between simulation and observations in the low- and intermediate-mass bins (with the intermediate-mass groups of \simbac at greater discordance with observations than those of \simba) but strong agreement in the high-mass bin---much better than \simba in the group cores.

For Ca, the overlap of the \simbac and observed abundance profiles again increases relative to Si and S in the two lower-mass bins. In particular, in the low-mass bin, the degree of improvement is similar to that for \simba; however, in the intermediate-mass bin, \simbac is more consistent than \simba with observations of Ca abundance at $R \lesssim 0.1R_{500}$ in both amplitude and slope, while \simba remains in better agreement at $R \gtrsim 0.1R_{500}$. Moreover, in the high-mass bin, \simbac is now in better agreement with observed profiles at $R \gtrsim 0.1R_{500}$. In this bin, while \simbac still better matches the flatter observed core abundance profiles of Ca than \simba, \simba better matches the core amplitude. Thus, the clear improvement from \simba to \simbac found for Si, S, and Fe in high-mass groups becomes muddled for Ca.

\subsubsection{\simbac--\simba Comparison}
\label{subsubsec:results_si_s_ca_simbac_simba_comparison}

The evolution of the abundance profiles with increasing group mass for Si, S, and Ca is commensurate with that for Fe. \simbac and \simba show similar profiles in the low-mass bin, which switches in the intermediate-mass bin to \simbac having steeper and higher-amplitude profiles, and then switches again to flatter central \simbac profiles in the high-mass bin. These results are consistent with the notion that \simbac retains its metals to a larger degree in the cores of high-mass groups compared to \simba, but does not do so in low-mass groups.

\section{Discussion}
\label{sec:discussion}

In this section, we qualitatively assess which simulation is most in agreement with observations, and investigate and discuss potential reasons for the differences and similarities we find between \simba and \simbac.

\subsection{Does SIMBA-C Improve on Agreement with Observations?}
\label{subsec:discussion_which_agrees_better?}

In terms of measuring and constraining resolved X-ray abundances with low uncertainty (i.e., high spatial and spectral resolution), observations are still relatively in their adolescence. Cosmological simulations, while advanced and successful in many ways, also have their limitations, e.g., low-resolution group cores, ad hoc models for stellar and AGN feedback, and, in our case, the neglect of metal diffusion (a process that is likely vital in the real universe for distributing metals). Many of these are being addressed with the advent of faster gravity and hydrodynamics codes (e.g., \textsc{swift}~\citep{schallerSwiftModernHighly2024}, which will be employed by \textsc{kiara}), high-resolution simulations (e.g., \textsc{hyenas}~\citep{cuiHYENASProjectPrediction2024a,jenningsHyenasXrayBubbles2025}, \textsc{Romulus}~\citep{tremmelRomulusCosmologicalSimulations2017,jungMassiveCentralGalaxies2022,saeedzadehCoolGustyChance2023}, \textsc{tng-cluster}~\citep{nelsonIntroducingTNGClusterSimulation2024,ayromlouAtlasGasMotions2024,truongXrayinferredKinematicsCore2024,lehleHeartGalaxyClusters2024,rohrCoolerIntraclusterMedium2025}), and increasingly realistic and physically motivated sub-grid models, like the AGN model of \textsc{obsidian}~\citep{rennehanObsidianModelThree2024}.

At the moment, though, asking which simulation is in better agreement with data may not always yield robust answers. In particular, metals like C, N, Ne, Ar, Cr, Mn, and Ni---elements that are especially difficult to pick out of the noise and thermal continuum in low-mass/low-temperature group spectra---observationally have low statistics, large scatter, and a lack of abundance measurements extending past group cores. However, as shown in Section~\ref{sec:results}, our results broadly demonstrate improvement from \simba to \simbac in the ability to predict metal abundances and their trends across a range of radii and~masses.

For C, N, O, Ne, and Mg, the metals that are produced primarily by AGB stars or core-collapse SNe, the \simbac abundance profiles have substantially lower amplitudes than \simba in the low-mass groups, generally bringing them into better agreement with observations. The high-mass group profiles are flatter in \simbac than in \simba, improving agreement between \simbac and observations. This strongly suggests that the upgrades implemented in \simbac have a positive impact on the mechanisms that bring these metals to and distribute them in the IGrM.

On the other hand, as noted, Si, S, Ca, and Fe, the metals most produced by SNe Ia of those that we consider, are too abundant in the IGrM of low-mass groups in both \simba and \simbac, to a similar degree. For these same metals, increasing $M_{500}$ by about 0.5 dex causes the \simba IGrM abundances in the range of $0.1 \lesssim R/R_{500} \lesssim 1$ to noticeably decrease, bringing them more in line with observations. \simbac, on the other hand, changes little across that mass range, and remains discrepant with observations. The reason for this is currently not clear, but may have something to do with differences in the distribution of metals with $L_X$ across IGrM gas elements between \simbac and \simba. As before, in the high-mass bin, it appears that \simba is confining these metals largely in the group cores, whereas \simbac spreads them out more uniformly, permitting it to better match observations. This is potentially a consequence of delayed SNIa enrichment being more sensitive to the (altered) details of feedback than prompt enrichment.

The \simbac Ni abundance profiles broadly match existing observations, and, where they do not, observations are limited with large scatter and/or uncertainty. Therefore, \simbac and the \texttt{Chem5} model hold promise for predicting the IGrM abundance distributions of many metals.

Lastly, observations display a flattening of the Fe abundance profiles in group cores across a range of masses (e.g., the X-COP sample~\citep{ghizzardiIronXCOPTracing2021a}, as shown in the study of Braspenning et~al. \cite{braspenningFLAMINGOProjectGalaxy2024}, and the samples of Sun \cite{sunHotGasGalaxy2012}, Sasaki et~al. \cite{sasakiMETALDISTRIBUTIONSOUT2014}, Mernier et~al. \cite{mernierAstrophysicsRadialMetal2017}, and Lovisari and Reiprich \cite{lovisariNonuniformityGalaxyCluster2019}, as plotted by Gastaldello et~al. \cite{gastaldelloMetalContentHot2021}). \simbac better reflects this feature than \simba, in particular at high group masses, and, as such, is an improvement.

Partial forward modeling of our simulated abundance profiles is key for accurately comparing simulations to observations. However, attempting to ascribe reasons for the different trends between \simba and \simbac is not necessarily meaningful, as the projection and $L_X$ weighting of abundances do not necessarily transparently reflect the underlying metal enrichment and distribution mechanisms. For example, the differences between \simba and \simbac may be over- or under-exaggerated, or skewed because metal abundance and gas X-ray luminosity do not strictly track each other (see, e.g., the study of Braspenning et~al.~\citep{braspenningFLAMINGOProjectGalaxy2024}, who compare X-ray luminosity, mass, and volume weighting). We leave this to Section~\ref{subsec:discussion_changes}.

It is pertinent to mention that other biases have been shown to be important to take into account when comparing simulated and observed abundance profiles, in particular, the CC/NCC dichotomy. In both simulations~\citep{fabjanSimulatingEffectActive2010,rasiaCoolCoreClusters2015,biffiHistoryChemicalEnrichment2017,vogelsbergerUniformityTimeinvarianceIntracluster2018,barnesClusterEAGLEProjectGlobal2017,barnesCensusCoolcoreGalaxy2018,braspenningFLAMINGOProjectGalaxy2024} and observations~\citep{leccardiRadialMetallicityProfiles2008,johnsonAbundanceProfilesCool2011,ettoriEvolutionSpatiallyResolved2015,mernierAstrophysicsRadialMetal2017,ghizzardiIronXCOPTracing2021a}, CC clusters tend to exhibit higher metallicity cores and more negative metallicity gradients than NCC clusters. To our knowledge, statistical studies of the correlation of core metallicity with CC/NCC status at the \emph{{group}} scale have not been conducted; however, a similar characteristic may exist.

Additionally, there is preferential inclusion of CC systems in X-ray-selected samples due to their dense, bright centers (see \citep{pearsonGalaxyMassAssembly2017,hendenFABLESimulationsFeedback2018}). The same biases may be responsible for at least part of the discrepancy between simulations and observations. While unknown currently for \simbac, the \simba $z=0$ galaxy group population consists of primarily NCC systems, possibly a result of the overly efficient feedback~\citep{oppenheimerSimulatingGroupsIntraGroup2021}. On the other hand, \textsc{CHEERS} comprises \emph{{only}} CC groups and three out of four groups in the Sarkar et~al. \cite{sarkarChemicalAbundancesOutskirts2022} sample are CC, thus potentially making the simulation--observation comparison not strictly ``like to~like''.

This merits a deeper investigation of CC/NCC classifications in \simba and \simbac and how it impacts their abundance profiles, as well as the incorporation of a wider range of groups into our comparison observational samples that better represent the true distribution of CC and NCC groups in the Universe. This is out of the scope of the present study, but will be explored in the future.

\subsection{Potential Reasons for Changes from SIMBA to SIMBA-C}
\label{subsec:discussion_changes}

The new \texttt{Chem5} model in \simbac is more complex than the instantaneous recycling of metals model of \simba. It is therefore difficult to trace changes in the results to their source. To attempt to decipher the primary drivers of these changes, we conduct some preliminary tests and refer to the recent results from \simbac~\citep{houghSIMBACUpdatedChemical2023a,houghSimbaCEvolutionThermal2024a}. Subsequent studies will more deeply explore the origins of these changes.

As such, we compute spherically averaged (3D) mass-weighted IGrM abundance profiles (Figure~\ref{fig:sim_3d_Zmass_solar_profiles_ylog_r500}) to investigate the differences between the two simulations, since these better represent the intrinsic abundance distributions than the projected $L_X$-weighted profiles. We include only those elements that are tracked in both simulations, and order them by atomic number. Additionally, we now extend the upper radial limit to $10R_{500}$, enabling us to assess the extent to which the different mass bins and two simulations converge in and beyond group outskirts.

\subsubsection{Abundance Profile Amplitudes}
\label{subsubsec:discussion_amplitudes}

\begin{figure}[H]
    \begin{adjustwidth}{-\extralength}{0cm}
    \centering
    \includegraphics[width=0.99\linewidth]{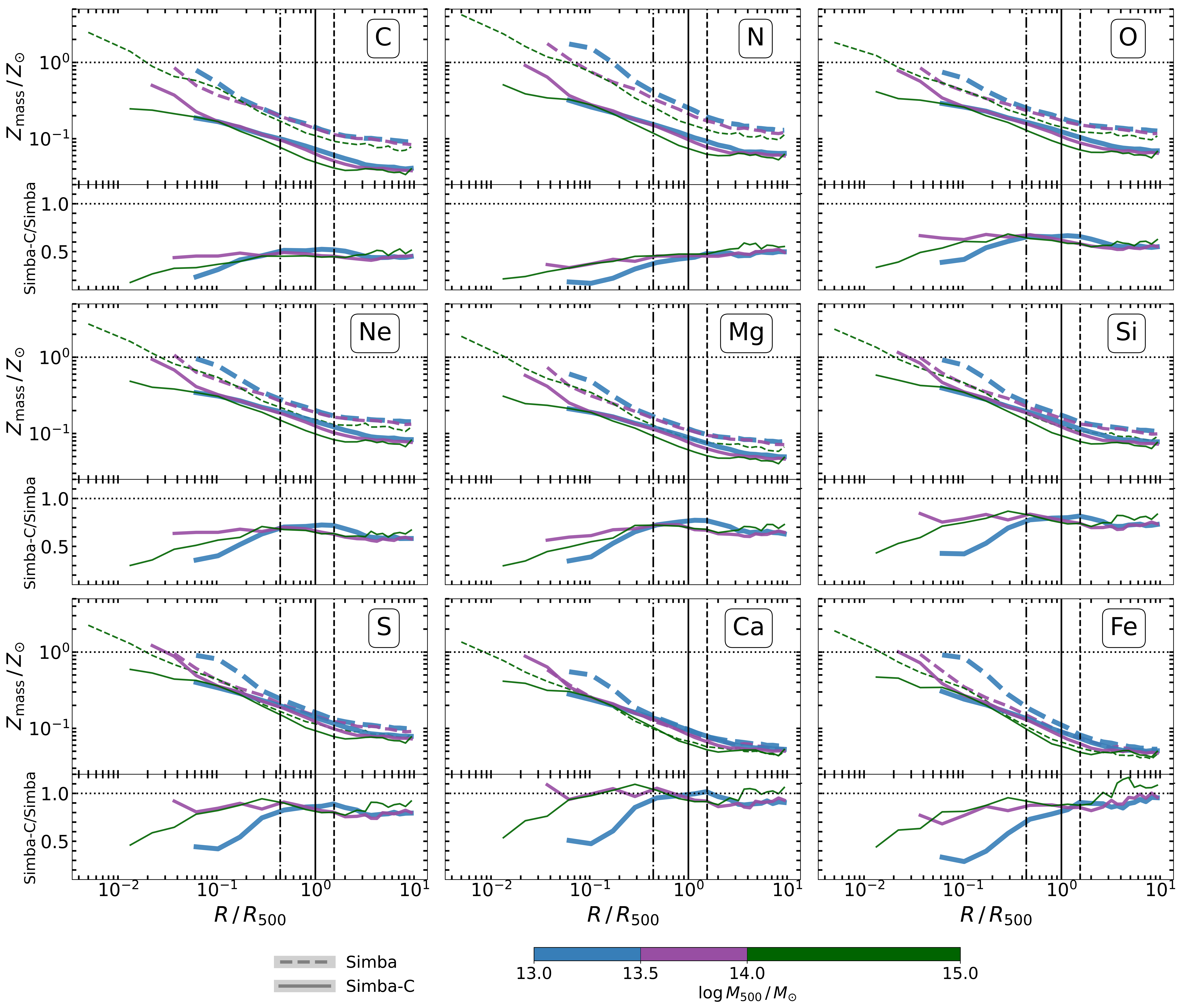}
    \end{adjustwidth}
    \caption{Median 3D mass-weighted logarithmically scaled metal abundance profiles for all metals in both \simba and \simbac, in our three previously defined $M_{500}$ bins. These profiles are scaled by their solar abundances from Asplund et~al. \cite{asplundChemicalCompositionSun2009}. The 16{th}--84{th} percentile regions are omitted for clarity; however, comparing \simbac to \simba profiles, they generally exhibit discordance at the level of $\gtrsim1\sigma$ at $R \lesssim R_{2500}$ and $\gg1\sigma$ at $R \gtrsim R_{2500}$ for C, N, O, Ne, and Mg; and $\sim$1$\sigma$ at $R \lesssim R_{2500}$ and $\gtrsim1\sigma$ at $R \gtrsim R_{2500}$ for Si, S, Ca, and Fe. The lower part of the subplot for each element shows the ratios of the \simbac and \simba profiles in each mass bin.}
    \label{fig:sim_3d_Zmass_solar_profiles_ylog_r500}
\end{figure}

Figure~\ref{fig:sim_3d_Zmass_solar_profiles_ylog_r500} indicates there are relatively minor changes in the outer region ($R \gtrsim R_{2500}$, up to $10R_{500}$) abundances of both \simba and \simbac groups as $M_{500}$ increases, with a slight but noticeable drop from the intermediate-mass bin (purple lines) to the high-mass bin (green lines), similar to the $L_X$-weighted profiles. On the other hand, as demonstrated by the ratio profiles, the abundances of \simbac are essentially never greater than those of \simba, contrary to what we find for the $L_X$-weighted profiles of Si, S, Ca, and Fe in the intermediate-mass bin. These results suggest that some trends found in the 2D $L_X$-weighted profiles may be strongly impacted by the emission weighting and/or projecting. In supplementary tests, we find that both emission weighting and projecting can contribute significantly to differences, although emission weighting tends to dominate.

In Section~\ref{sec:results}, we found that, for most metals (excluding C, N, and Ni), both \simba and \simbac over-enrich the outer regions of the IGrM ($R\gtrsim0.1R_{500}$) in low-mass groups. Braspenning et~al. \cite{braspenningFLAMINGOProjectGalaxy2024} attribute a similar discrepancy in the \textsc{flamingo} simulations to unduly high assumed nucleosynthetic yields, or to the total stellar masses being too high in the simulations and over-producing metals. As with the $L_X$-weighted profiles, Figure~\ref{fig:sim_3d_Zmass_solar_profiles_ylog_r500} demonstrates that the lighter elements, like C and N, exhibit noticeably lower abundance profile amplitudes in \simbac than in \simba in all mass bins by up to $\sim 0.5$ dex. This is likely a consequence of the new yields in \texttt{Chem5} being lower overall than those in \simba.

\vspace{-4pt}

Another possible reason contributing to this is as follows: Via the instantaneous recycling of metals in \simba, gas elements could self-enrich before forming stars through SNII and prompt SNIa feedback, with further enrichment from nearby stars (i.e., due to delayed feedback if the gas element in question is one of the 16 nearest neighbors to a star particle) and metal-loaded stellar winds~\citep{daveMUFASAGalaxyFormation2016,daveSimbaCosmologicalSimulations2019}. This may result in highly enriched gas escaping galaxies to join the IGrM. In contrast, \simbac eliminates self-enrichment and metal-loaded winds, with metallicity only increasing via nearby star particles' feedback. Consequently, stars and gas (both ISM and IGrM) in \simbac are expected to be less enriched. After the implementation of \texttt{Chem5}, Hough et~al.~\citep{houghSIMBACUpdatedChemical2023a} performed recalibrations on \simbac to improve low-redshift metal production and the agreement of the simulated and observed GSMFs. Optimizing between these two resulted in lower ISM~\citep{houghSIMBACUpdatedChemical2023a} and IGrM~\citep{houghSimbaCEvolutionThermal2024a} gas-phase metallicities in \simbac than in \simba, by up to $\sim$0.2 dex.

That said, as atomic number increases, the IGrM abundance profiles of \simba and \simbac approach each other. This culminates with Ca and Fe, for which there is significant overlap between the two simulations, especially in the group outskirts. Thus, for these metals, there is no reduction in the abundance profile amplitudes that would enable \simbac to better match low-mass group observations than \simba. We postulate this may be a consequence of the inclusion of HNe in \texttt{Chem5}, which is not included in \simba. HNe produce a greater amount of Fe and $\alpha$ elements than the lower-energy SNe II~\citep{kobayashiOriginElementsCarbon2020,houghSIMBACUpdatedChemical2023a}. On the other hand, for metals including and heavier than O, there is a net reduction of SNe II yields from \simba to \simbac due to the inclusion of ‘failed’ SNe yields. It is possible that the ``extra'' HNe products make up the difference between the \simba yields and the lower \simbac yields of the heavier metals.

These points highlight systematic issues with sub-grid models at the group scale. The pervasive nature of this issue from \simba to \simbac, at least for the heavier metals, hints at the source being a sub-grid model that has been minimally altered or not changed in the upgrade to \simbac, such as AGN feedback, the stellar-wind-powered dispersal of metals, radiative cooling and heating, or the lack of metal diffusion.

Along these same lines, the nature of feedback processes and their interaction with group environments may play a significant role. Lower-mass groups, with their shallower gravitational potentials, are thought to be more easily impacted by AGN or stellar feedback~\citep{oppenheimerSimulatingGroupsIntraGroup2021}. Even though components of the AGN feedback model were altered from \simba to \simbac, the overall model did not change. Similarly, although there was an overhaul in the stellar feedback model, the general treatment of stellar winds remained unchanged. Therefore, it is plausible that, at the low-mass group scale, (inappropriately modeled) feedback is important for setting the metal distribution.

Hough et~al. \cite{houghSIMBACUpdatedChemical2023a} find sizable differences going from \simba to \simbac in both their galactic gas-phase and stellar abundances and abundance ratios, which lead to improvements with respect to observations. This may imply that, although the enrichment within galaxies is improved in \simbac relative to \simba, the mechanisms that bring metals out of the ISM phases and disperse them into the IGrM phase (in temperature and location), such as stellar winds and turbulent mixing, remain inadequately modeled at the \mbox{group scale}.

To this point, Rennehan et~al. \cite{rennehanDynamicLocalizedTurbulent2019} present and test a new method for dynamically calculating turbulent diffusivity. Subsequently, Rennehan \cite{rennehanMixingMatters2021} implements an anisotropic eddy viscosity and metal-mixing model into a Lagrangian astrophysical simulation. Most relevantly, they show that the hot diffuse gas around individual galaxies (known as the circumgalactic medium, or CGM) is particularly sensitive to the treatment of turbulent diffusion, and their models result in earlier metal mixing and more lower-metallicity gas in the metal distribution functions of the CGM than other models, including the oft-employed constant coefficient Smagorinsky model. As discussed, both \simba and \simbac produce over-enriched low-mass groups for most metals. Extending the results of Rennehan et~al.~\cite{rennehanDynamicLocalizedTurbulent2019} and Rennehan~\cite{rennehanMixingMatters2021} from the CGM to the low-mass IGrM suggests that explicit and realistic sub-grid modeling of metal diffusion could help remedy this. {Further, other recent studies find that turbulence may be important for generally producing realistic galaxies in simulations, in particular in the ISM, as well as generating realistic turbulent properties of the IGrM (e.g.,~\citep{grothTurbulentPressureSupport2024,sotiraImpactAGNFeedback2024,kempfNonlinearSaturationEnergy2024,beattieMagnetizedCompressibleTurbulence2024,beattieLongKolmogorovForward2025}}).

\subsubsection{Abundance Profile Slopes}
\label{subsubsec:discussion_slopes}

The shapes of abundance profiles are noteworthy, often considered to impart information about the degree of mixing in the IGrM and when enrichment occurs. Although neither cosmological simulations nor observations are at the level of precision necessary for comparing them on equal footing, it is of interest to start to take a look at this.

\textls[-11]{In Figure~\ref{fig:sim_3d_Zmass_solar_profiles_ylog_r500}, similar to the $L_X$-weighted abundance profiles, the \simbac mass-weighted profiles are generally less steep in the group inner regions than those of \simba. Further, the abundance profiles of all metals in all mass bins in both \simba and \simbac exhibit flattening at $R \gtrsim R_{200}$. Resolved X-ray observations of \emph{cluster} ($M_{500} \gtrsim 10^{14} \, \mathrm{M_\odot}$) outskirts also find flat metallicity profiles (e.g.,~\citep{wernerUniformMetalDistribution2013,urbanUniformMetallicityOutskirts2017,ghizzardiIronXCOPTracing2021a}). At the \emph{group} scale ($M_{500} \lesssim 10^{14} \, \mathrm{M_\odot}$), observations are less robust and more varied. Some lower-mass group samples, like those of Sarkar et~al. \cite{sarkarChemicalAbundancesOutskirts2022}, exhibit relatively flat outskirt profiles, while others, like those of Rasmussen and Ponman~\cite{rasmussenTemperatureAbundanceProfiles2007} and Mernier et~al. \cite{mernierAstrophysicsRadialMetal2017}, cannot formally constrain the outskirts\endnote{According to Mernier et~al. \cite{mernierAstrophysicsRadialMetal2017}, neither flat nor decreasing outskirt profiles are favored over the other in \textsc{CHEERS}. On the other hand, the Sarkar et~al. \cite{sarkarChemicalAbundancesOutskirts2022} sample contains more robust outskirt abundance profiles due to \emph{Suzaku}'s stable particle background and high spectral sensitivity below 1 keV~\citep{koyamaXRayImagingSpectrometer2007}. This latter sample, however, is quite small.}.}

The \emph{XRISM}/Resolve X-ray spectrometer, with the highest spectral resolution of any current similar instrument, has the ability to reduce abundance uncertainties significantly and improve observations of different radial regions in groups and clusters~\citep{ishisakiResolveInstrumentXray2018}. In the near future, the \emph{Athena} X-ray Integral Field Unit spatially resolved spectrometer~\citep{barretAthenaXrayIntegral2023} and \emph{AXIS}~\citep{mushotzkyAdvancedXrayImaging2019a,reynoldsOverviewAdvancedXray2023} will further improve on this, {the former obtaining highly accurate abundance profiles through high spectral and spatial resolutions and the latter measuring ICM abundances in various spatial regions with high sensitivity}. Without these facilities, statistically confirming the outskirt abundance profile slopes in observations of low-mass groups---and comparing them with simulations like \simba and \simbac---would be difficult given the current technological limitations.

Some studies find that the production of $z=0$ metallicity profiles with flat outskirts in zoom cosmological simulations of clusters requires an ``early enrichment scenario''\linebreak \mbox{(e.g.,~\citep{sijackiHydrodynamicalSimulationsCluster2006,fabjanSimulatingEffectActive2010,biffiHistoryChemicalEnrichment2017,biffiOriginICMEnrichment2018}).} Specifically, early ($z \gtrsim 2-3$) AGN feedback is needed to sufficiently expel enriched gas from galaxies into the IGrM and quench star formation, keeping metals in the hot diffuse phase and preventing them from being locked back in the stellar phase~\citep{fabjanSimulatingEffectActive2010}.

AGN feedback is found to be more efficient than stellar feedback in distributing metal-rich gas to large cluster-centric distances in small high-$z$ halos (prior to the assembly of $z=0$ groups and clusters), reducing the build-up of high-metallicity cores and increasing outskirt abundances~\citep{biffiHistoryChemicalEnrichment2017,biffiOriginICMEnrichment2018}. Gas dynamical processes---associated with the merging of these halos and continued energy output from feedback---promote the mixing of the pre-enriched and hot diffuse gas into relatively uniform distributions~\citep{fabjanSimulatingEffectActive2010}.

It is important to note that, in all of these studies, AGN feedback is not \emph{{required}} to disperse metals from galaxies out to the CGM or IGrM in their cosmological simulations. Those without AGN feedback can still exhibit low-to-moderate abundances in galaxy, group, or cluster outskirts when only accounting for star formation and stellar feedback. In fact, high-resolution simulations that track the evolution of high-velocity nuclear outflows (e.g.,~\citep{faucher-giguerePhysicsGalacticWinds2012,gaborActiveGalacticNucleidriven2014}) demonstrate that these outflows transition into an expanding wind of shocked gas before escaping the central region of galaxies. When they subsequently encounter the galactic disk, they follow the path of least resistance, resulting in their preferential expulsion perpendicular to the disk. As such, the metal-rich ISM is found to be affected only to a limited degree by AGN outflows. However, the cost of removing the AGN is the production of overly star-forming BGGs with $\sim$2--4$\times$ too many stars (e.g.,~\citep{mccarthyCaseAGNFeedback2010}).

Critically, most of these simulations, including \simba and \simbac, do not include any explicit implementation of metal mixing, a process potentially crucial for setting metal distributions in the real IGrM. Rennehan et~al. \cite{rennehanDynamicLocalizedTurbulent2019} and Rennehan \cite{rennehanMixingMatters2021} demonstrate that improved models of turbulent diffusion result in the rapid mixing of fluid properties, including metals. This reveals that, not only can the implementation of metal mixing have an impact on the timing of enrichment, but also that observables (like abundance profile shapes) are not necessarily good indicators of the time or efficiency of mixing because metals may be able to mix uniformly on short timescales regardless of when enrichment~occurs.

Additionally, \simba and \simbac employ hydrodynamically decoupled winds and jets for AGN feedback, meaning AGN feedback should not be capable of bringing metal-enriched gas from the ISM of the central galaxy out to larger radii (unless the gas located near the SMBH is enriched). Other mechanisms, like mergers or stellar feedback with metal-enriched winds, are required. Once the enriched gas particles are at sufficiently large radii, where the AGN wind and jet particles can recouple, AGN feedback could then ostensibly redistribute the metal-enriched gas. Therefore, a complex blend of physical processes, such as cooling, turbulence, star formation, and AGN and stellar feedback (most of which are not well modeled in cosmological simulations), as well as potentially simulation resolution, is likely responsible for setting the particular distribution of metals in the IGrM.

\subsubsection{IGrM Mass and Metal Mass Profiles}
\label{subsubsec:discussion_mass_profiles}

\begin{figure}[H]
    \begin{adjustwidth}{-\extralength}{0cm}
    \centering
    \includegraphics[width=0.99\linewidth]{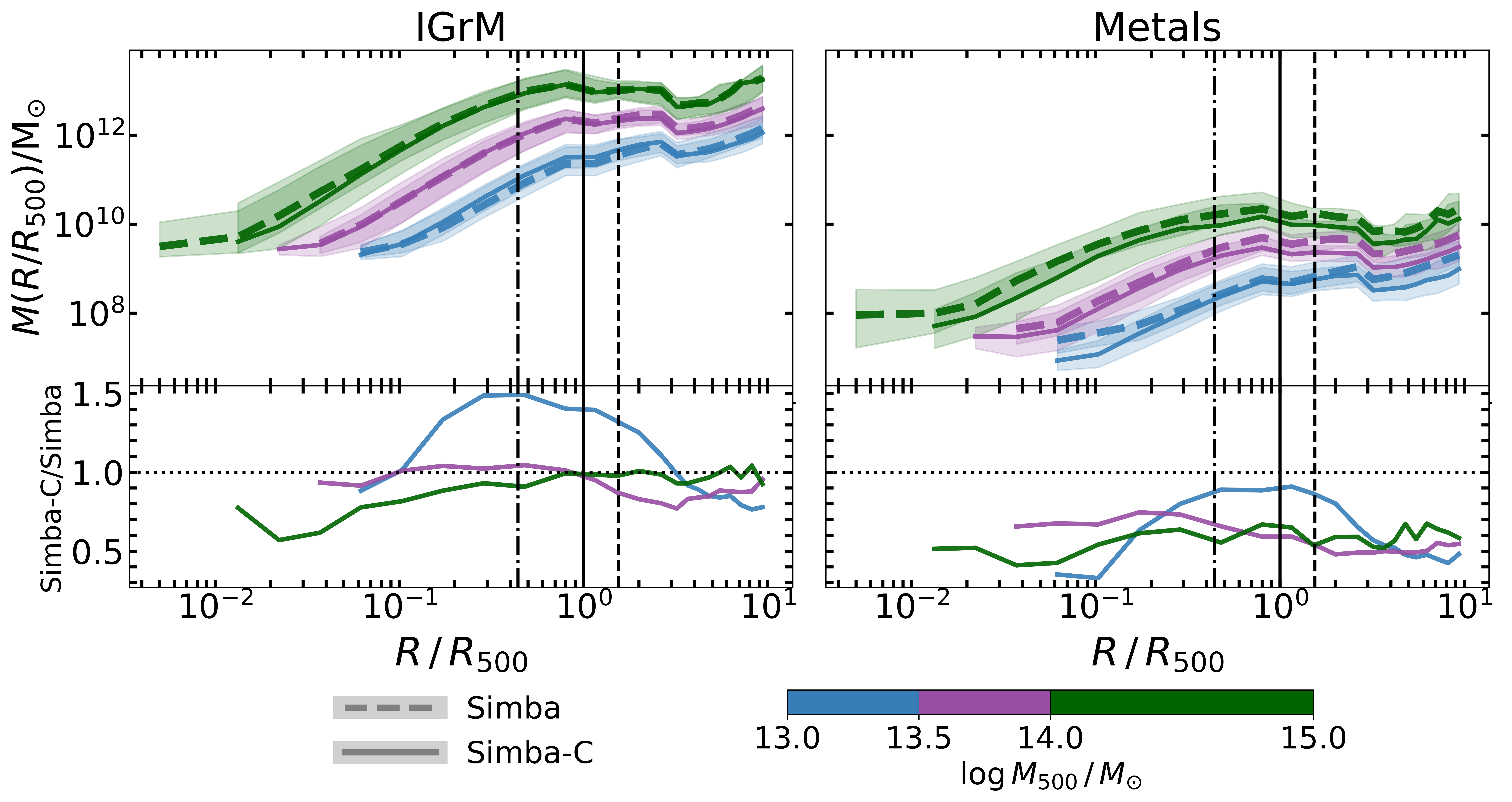}
    \end{adjustwidth}
    \caption{Median mass profiles of the IGrM (\emph{left}) and the sum total of IGrM-phase metals (\emph{right}) in \simba and \simbac, in our three $M_{500}$ bins. Shaded regions represent the 16{th}--84{th} percentiles of the scatter. The lower half of each subplot shows the ratios of the \simbac and \simba profiles for each mass bin.}
    \label{fig:sim_igrm_metals_3d_mass_profiles_ylog_r500_compare_bins}
\end{figure}

Next, in Figure~\ref{fig:sim_igrm_metals_3d_mass_profiles_ylog_r500_compare_bins}, we investigate the separate components that form the abundance profiles, namely, the total mass profile of IGrM-phase metals (summed together) and the mass profile of all gas in the IGrM. We find that, in the low-mass bin (blue lines), the IGrM mass profile increases from \simba to \simbac at most radii. Looking at the intermediate-mass bin (purple lines), we note that there is no increase in IGrM mass within $R_{500}$ and, instead, a slight decrease at larger radii. The high-mass bin (green lines) yields the opposite, i.e., a decrease within $R_{500}$ and no change outside $R_{500}$.

\vspace{-3pt}
The dependence on the halo mass of the $M_\mathrm{IGrM}$ \simbac--\simba ratio profiles, as well as the two simulations' identical cosmologies, indicates that alterations to the feedback models in \simbac are plausible candidates for the origin of these differences. The altered stellar wind feedback likely contributes, in particular, the reduction in the stellar wind velocity scaling $a$ in \simbac by a factor of $\sim$2 compared to \simba. As previously noted, the gas in lower-mass groups is more prone to feedback and other baryonic processes than that in higher-mass groups~\citep{oppenheimerSimulatingGroupsIntraGroup2021}. This implies that a change in feedback strength will be more noticeable in the gas properties of these lower-mass groups. Specifically, a lower fraction of gas may be blown to larger radii (or even out of the groups altogether), allowing them to retain more IGrM mass. Hence, we observe an increase in IGrM mass in the low-$M_{500}$ bin, but more limited changes in the higher-mass bins.

For $R \lesssim R_{500}$, the low-mass bin shows an overall increase in IGrM mass from \simba to \simbac, while the intermediate- and high-mass bins show minimal change (in the high-mass bin, the decrease is largely confined to the group cores, which contribute little to the total mass). This is consistent with the findings of Hough et~al. \cite{houghSimbaCEvolutionThermal2024a}, who find a higher IGrM mass fraction ($M_{\mathrm{IGrM},500}/M_{500}$) in \simbac than in \simba for halos in the range $10^{13} \lesssim M_{500}/\mathrm{M_\odot} \lesssim 10^{13.5}$ but negligible differences at higher masses.

Moreover, Khrykin et~al. \cite{khrykinCosmicBaryonPartition2024} show that, in \simba, the jet mode AGN feedback is key in distributing baryons between the CGM and IGM at $z$~=~0--1, increasing the IGM baryon fraction by $\sim$$20$\% relative to \textsc{simba nojet}. They also find that AGN feedback is the most effective mechanism for this redistribution in halos with masses $10^{12} \lesssim M_{200}/\mathrm{M_\odot} \lesssim 10^{14}$ (approximately $5 \times 10^{11} \lesssim M_{500}/\mathrm{M_\odot} \lesssim 5 \times 10^{13}$), but that the baryon distribution in halos outside this mass range is only weakly sensitive to the specific feedback prescription. This is in very good agreement with our results, which also exhibit significant changes in the IGrM mass profiles of similarly low-mass groups but minor changes above this mass range.

Relative to \simba, we find that IGrM metal mass decreases by $\sim$10--70\% across all radii and halo masses in \simbac. The reduced metal content leads to the lower mass-weighted abundances in Figure~\ref{fig:sim_3d_Zmass_solar_profiles_ylog_r500}, since the IGrM masses do not simultaneously decrease. As before, this decrease may reflect the overall reduced nucleosynthetic yields and the lack of self-enrichment in \texttt{Chem5}.

It is possible that the lower metal cooling rates in \simbac (due to reduced abundances) keep more gas in the hot diffuse phase of the IGrM, which could help explain the IGrM gas mass increase from \simba to \simbac for low-mass groups~\citep{houghSIMBACUpdatedChemical2023a,houghSimbaCEvolutionThermal2024a}. However, we note that, Fe being the primary coolant for the hot diffuse gas (at least around its virial temperature), and there being minimal differences between the \simba and \simbac mass-weighted Fe abundance profiles in Figure~\ref{fig:sim_3d_Zmass_solar_profiles_ylog_r500}, it is unlikely that these changes can fully account for the observed differences in the IGrM mass profiles.

More feasibly, the alterations to the AGN and stellar feedback models impacted the timing of mass assembly in the groups, resulting in a build-up of hot diffuse gas in low-mass groups. In \simbac, for example, the reduced stellar wind velocity normalization and the effectively delayed AGN jet activation may allow the group gravitational potentials to deepen rapidly prior to feedback being able to blow gas out of the groups. In \simba, stellar feedback may be too strong or AGN jets may activate too early, blowing gas out and leaving little behind in the IGrM. Hough et~al. \cite{houghSimbaCEvolutionThermal2024a} find quite low IGrM mass fractions in low-mass groups, particularly in \simba.

\begin{figure}[H]
    \includegraphics[width=0.75\linewidth,keepaspectratio]{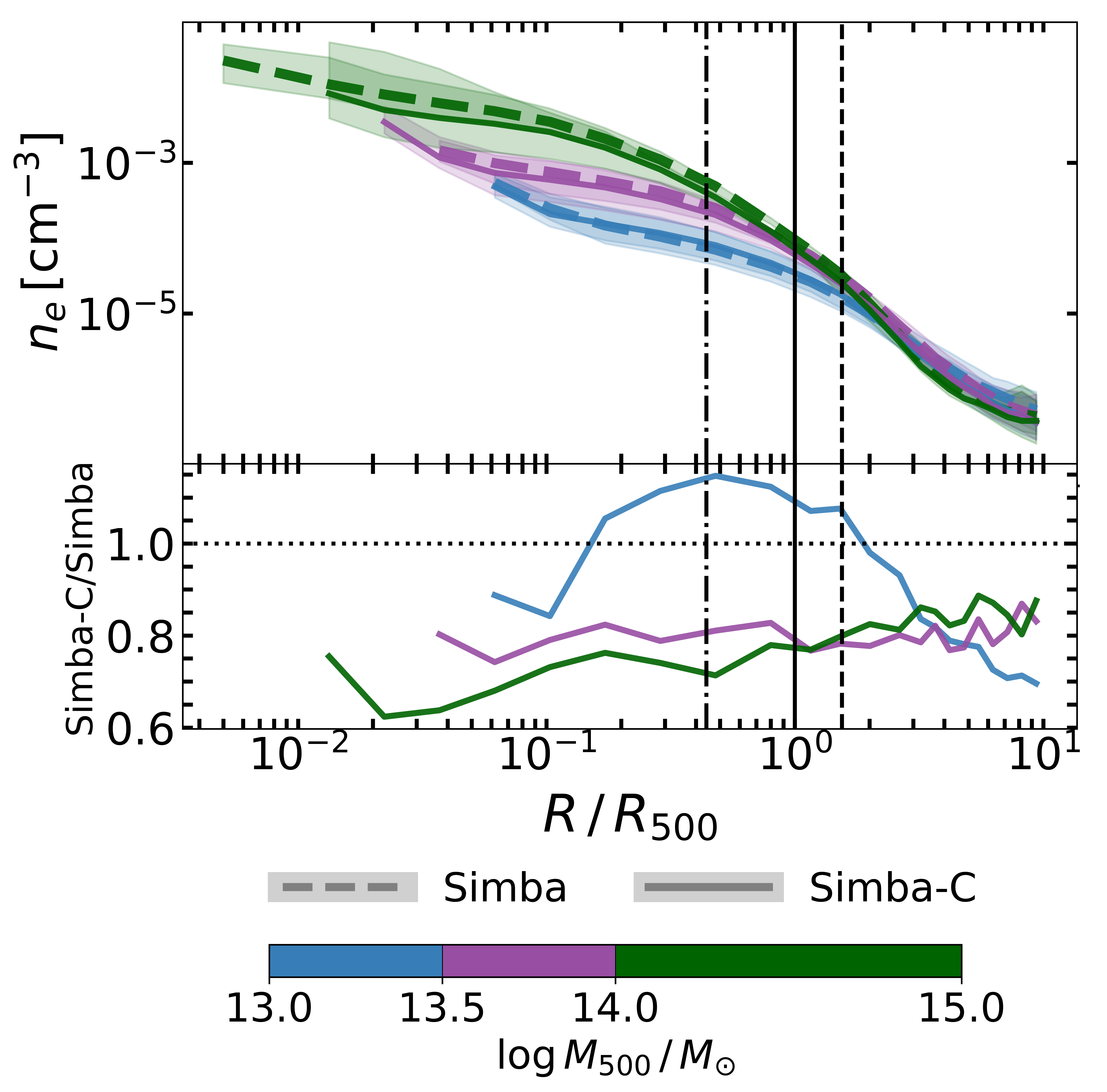}
    \caption{Median electron number density profiles of the IGrM in \simba and \simbac, in our three $M_{500}$ bins. Shaded regions represent the 16{th}--84{th} percentiles of the scatter. The lower half of the figure shows the ratios of the \simbac and \simba profiles for each mass bin.}
    \label{fig:sim_ne_profiles}
\end{figure}

{Further, changes in the IGrM gas density may alter the IGrM mass. In Figure~\ref{fig:sim_ne_profiles}, we find that \simbac produces IGrM gas density profiles that are $\sim$25\% lower in amplitude than \simba in the intermediate- and high-mass bins. On the other hand, it is more variable in the low-mass bin; at $R \lesssim 0.2R_{500}$ and $R \gtrsim R_{200}$, the gas density is slightly reduced in \simbac, whereas, at $0.2R_{500} \lesssim R \lesssim R_{200}$, there is a slight increase.}

{The volumetric gas cooling rate is proportional to $n_e$ and $Z$ as $\dot{E} \propto n_e^2 \Lambda(Z,T)$, where $\Lambda(Z,T)$ is the cooling function. Smaller differences in $n_e$ become more important, while changes in $Z$ are less so because $\dot{E}$ is only linearly proportional to it. The rate of cooling contributes heavily to the evolution and state of the IGrM, including its total mass and mass profile. Increased densities enhance the cooling rate, drawing more gas out of the hot diffuse phase, and vice versa. However, in this case, differences in gas density between \simba and \simbac are likely sub-dominant to the effects of the altered AGN and stellar sub-grid feedback models because the changes in density are not substantial.}

Putting this all together, it appears that the lower-amplitude mass-weighted abundance profiles in high-mass \simbac groups (mostly seen with C, N, O, Ne, and Mg) are primarily a consequence of fewer metals in the IGrM in \simbac. On the other hand, the same trend in low-mass \simbac groups is additionally driven by an increased IGrM gas~mass.

\section{Conclusions}
\label{sec:conclusions}

We conclude that the integration of the \texttt{Chem5} chemical enrichment model in \simba substantially impacts the abundance and distribution of metals in the IGrM, broadly improving their match to observations across two orders of magnitude in terms of group mass and a wide array of X-ray-relevant elements. \simbac, to the best of our knowledge, makes use of the most realistic chemical enrichment model (\texttt{Chem5}) currently feasible to implement in a cosmological simulation. It has also been tested against galaxy-scale properties~\citep{houghSIMBACUpdatedChemical2023a}, global X-ray observations of the IGrM~\citep{houghSimbaCEvolutionThermal2024a}, and the distribution of metals in the IGrM (this work), allowing its effects on galaxy formation and evolution to be well characterized across orders of magnitude in scale.

Our main results are as follow:

\begin{enumerate}
    \item On average, \simbac produces lower-amplitude projected $L_X$-weighted abundance profiles than \simba, with flatter profiles in the inner regions, particularly in high-mass groups but also in low-mass groups for some metals. These characteristics tend to be in better agreement with observations. The scatter within and across observational samples reduces their constraining power and sometimes allows both simulated samples to be consistent with observations. Additionally, in a few instances, \simba matches observations better than \simbac. For example, in our intermediate-mass bin, the Fe, Si, and S profiles of \simbac have greater amplitudes than those of \simba, resulting in its reduced agreement with observations.
    \item Despite the upgrade to \simbac, the low-mass groups ($13 < \log(M_{500}/M_\odot) < 14$) are generally in the worst agreement with observations, while the high-mass groups/clusters ($14 < \log(M_{500}/M_\odot) < 15$) are in the best agreement. This may be a consequence of the increasing importance in the shallower gravitational potential wells of the inadequate modeling of feedback processes, like AGN feedback or stellar winds, or the lack of any explicit implementation of critical physical mechanisms, like metal mixing. Therefore, issues in our cosmological simulations at the group scale continue to persist, and have yet to be addressed.
    \item {Specifically, \simbac is more consistent with IGrM observations of the abundances of most investigated metals (C, N, O, Ne, Mg, Si, S, Fe) at higher masses, especially in the group cores. \simbac further improves on \simba in its agreement with observed abundances of N (and, to some extent, C) across all group masses, suggesting the upgraded treatment of AGB star feedback in \textsc{Chem5} enhances the realism of the enrichment of the IGrM. On the other hand, \simba and \simbac both exhibit substantial discrepancy with IGrM observations of, for example, Si, S, Ca, and Fe at lower group masses, potentially indicating that further work on the understanding and modeling of core-collapse SN and SNIa feedback and enrichment is required.}
    \item Some of the features of the 2D emission-weighted abundance profiles are preserved in the 3D mass-weighted profiles, like lower amplitudes and flatter cores in \simbac than in \simba, while others are not, such as the aforementioned occasional higher amplitudes in \simbac compared to \simba. Additionally, emission weighting results in greater apparent abundances. These results suggest that spectroscopically inferred group abundance profiles are expressive to a limited degree of the underlying physical enrichment mechanisms, and care should be taken in their interpretation.
    \item The intrinsically lower IGrM abundances in \simbac than in \simba are largely attributed to reduced metal yields and the lack of instantaneous recycling in \texttt{Chem5} compared to \simba's enrichment model, resulting in a lower mass of IGrM-phase metals in \simbac. On the other hand, the increased IGrM mass in low-mass \simbac groups may be a consequence of the reduced stellar wind velocity normalization and the ``delayed'' AGN jet activation in \simbac, causing changes in the groups' mass assembly histories.
    \item We find that the \simba and \simbac mass-weighted abundance profiles draw closer together with increasing atomic number. We connect this to the addition of HNe enrichment in the \texttt{Chem5} model, which increases the production of Fe and $\alpha$ elements (e.g., Si, S, Ca), enhancing their abundance profiles to ``re-match'' those of \simba by approximately equaling the aforementioned overall decrease in metals in \simbac.
\end{enumerate}

As demonstrated by Hough et~al. \cite{houghSIMBACUpdatedChemical2023a}, \simbac shows improvements in matching observed galaxy properties, such as the galaxy stellar mass function and the stellar mass--metallicity relation, compared to the original \simba simulation. Hough et~al. \cite{houghSimbaCEvolutionThermal2024a} and the present study reveal that \simbac also improves on both global and detailed properties of the IGrM at the group scale. This indicates progress in directly addressing an enduring challenge in cosmological simulations: accurately and simultaneously reproducing galaxy-scale and group-scale observables.

The results of \simbac at the group scale additionally confirm the necessity of greater constraining power from observations to distinguish between different models of cosmic chemical enrichment, in particular at lower halo masses and for harder-to-observe metals in X-ray like C, N, Ne, Ca, and Ni. Current and upcoming X-ray missions, such as \mbox{\emph{XRISM}~\citep{teamScienceXrayImaging2020,tashiroXRISMXrayImaging2022}}, \emph{Athena}~\citep{barconsAthenaAdvancedTelescope2012,barretAthenaFirstDeep2013a,barretAthenaSpaceXray2020}, \emph{AXIS}~\citep{mushotzkyAdvancedXrayImaging2019a,reynoldsOverviewAdvancedXray2023}, and \emph{Lynx}~\citep{gaskinLynxXRayObservatory2019,schwartzLynxXrayObservatory2019}, aim to close this gap through increased sensitivity and spectral and spatial resolutions. These technological improvements will reduce the uncertainty of abundance and temperature measurements, enable the mapping of the IGrM out to larger radii, pick out the X-ray faint IGrM, make group and cluster X-ray samples more complete, and remove substantial observational biases. Concurrently, \simbac and future iterations will help guide observations regarding in which regions of the IGrM certain metals are likely to be found and in what abundances.

With the continuing progression in computational resources, algorithms, and numerical methods---as well as improved observing capabilities---sub-grid models must keep up. Realistic and self-consistent modeling of physical processes, such as AGN and stellar feedback, appears critical to reproducing observed enrichment of the hot diffuse gas in group environments. We therefore require not just accurate and physically motivated sub-grid models, like \texttt{Chem5}, but also improved treatment of key processes like turbulent diffusion~\citep{rennehanDynamicLocalizedTurbulent2019,rennehanMixingMatters2021} to be implemented in the next generation of cosmological simulations.



\vspace{6pt} 





\authorcontributions{Conceptualization, A.P.-B. and A.B.; data curation, A.P.-B. and F.M.; formal analysis, A.P.-B.; funding acquisition, A.P.-B. and A.B.; investigation, A.P.-B.; methodology, A.P.-B., Z.S., R.T.H., A.B. and W.C.; project administration, A.P.-B.; resources, D.R., A.B. and W.C.; software, A.P.-B., Z.S., R.T.H., D.R., R.D., C.K. and W.C.; supervision, A.B.; validation, A.P.-B.; visualization, A.P.-B.; writing---original draft, A.P.-B.; writing---review and editing, Z.S., R.T.H., D.R., R.B., V.S., A.B., R.D., C.K., W.C., F.M. and G.G. All authors have read and agreed to the published version of the~manuscript.}

\funding{This research was funded by a Natural Sciences and Engineering Research Council of Canada (NSERC) Canada Graduate Scholarship---Masters (CGS-M), grant number 585656/2023.}

\dataavailability{\textls[-11]{The \simbac simulation data underlying this article are publicly available at the Flatiron Institute upon reasonable request to Douglas Rennehan ({drennehan@flatironinstitute.org}). The published \simba simulations are available in the \simba simulation repository {at}  \url{http://Simba.roe.ac.uk/} (accessed on 1 January 2023).}}

\acknowledgments{The simulations and analyses reported in this article were enabled by HPC resources provided by the Digital Research Alliance of Canada (\url{alliancecan.ca}) award to A.B., specifically SciNet and the Niagara computing cluster. The \simbac simulation was run on the Flatiron Institute's research computing facilities (the Iron compute cluster), supported by the Simons Foundation. A.P.B. also acknowledges the \simba collaboration for the use of their simulations. A.B. acknowledges the support of the Natural Sciences and Engineering Research Council of Canada (NSERC) through its Discovery Grant program. A.B. also acknowledges support from the Infosys Foundation via an endowed Infosys Visiting Chair Professorship at the Indian Institute of Science and from the Leverhulme Trust via the Leverhulme Visiting Professorship at the University of Edinburgh. W.C. is supported by the Atracci\'on de Talento contract no. 2020-T1/TIC19882 granted by the Comunidad de Madrid in Spain, and the science research grants from the China Manned Space Project. He also thanks the Ministerio de Ciencia e Innovaci\'on (Spain) for financial support under project grant PID2021-122603NB-C21 and HORIZON EUROPE Marie Sklodowska-Curie Actions for supporting the LACEGAL-III project with grant number 101086388.
C.K. acknowledges funding from the UK Science and Technology Facility Council through grant ST/Y001443/1. {For observational comparisons with \textsc{CHEERS}, the material is based upon work supported by NASA under award number 80GSFC24M0006.} We thank A. Sarkar for directly contributing observational data. Finally, we acknowledge the l\rotatebox[origin=c]{180}{e}\'k$^{\mathrm{w}}$\rotatebox[origin=c]{180}{e}\textipa{\ng}\rotatebox[origin=c]{180}{e}n peoples on whose traditional territory the University of Victoria stands, and the Songhees, Equimalt, and WS\'ANE\'C peoples whose historical relationships with the land continue to this day.}

\conflictsofinterest{The authors declare no conflicts of interest. The funders had no role in the design of the study; in the collection, analyses, or interpretation of data; in the writing of the manuscript; or in the decision to publish the results.}





\appendixtitles{yes} 
\appendixstart
\appendix




\begin{adjustwidth}{-\extralength}{0cm}

\printendnotes[custom]

\reftitle{References}

\PublishersNote{}
\end{adjustwidth}
\end{document}